%% file: main.tex
\documentclass[10pt,twocolumn,letterpaper]{article}

\usepackage{cvpr}      % To produce the REVIEW version
\usepackage{multirow}
\usepackage[accsupp]{axessibility}  % Improves PDF readability for those with disabilities.
\input{preamble}
\definecolor{cvprblue}{rgb}{0.21,0.49,0.74}
\usepackage[pagebackref,breaklinks,colorlinks,allcolors=cvprblue]{hyperref}

\def\paperID{} % *** Enter the Paper ID here
\def\confName{CVPR}
\def\confYear{2026}

\title{\ours: Spatial-Physical REasoning via geometry Aware Diffusion}

\author{
Minzhang Li$^{*}$\quad
Kuixiang Shao$^{*}$\quad
Xuebing Li\quad
Yuyang Jiao\quad
Yinuo Bai\\
Hengan Zhou\quad
Sixian Shen\quad
Jiayuan Gu$^{\dagger}$\quad
Jingyi Yu$^{\dagger}$\\[0.5em]
ShanghaiTech University\\
{\tt\small \{limzh2022, shaokx2025, lixb2025, jiaoyy2022, baiyn2022,}\\
{\tt\small zhouha2025, shensx2024, gujy1, yujingyi\}@shanghaitech.edu.cn}\\[0.3em]
{\small $^{*}$Equal contribution.\quad $^{\dagger}$Corresponding authors.}
}

\begin{document}
\maketitle
\input{sec/0_abstract}    
\input{sec/1_intro}

\input{sec/2_related_work}
\input{sec/3_method}
\input{sec/4_experiments}
\input{sec/5_conclusion}

\section*{Acknowledgements}
This work was supported in part by the National Natural Science Foundation of China under Grant W2431046, National Key R\&D Program of China 2025YFA1309603, Central Guided Local Science and Technology Foundation of China YDZX20253100001001, and by MoE Key Lab of Intelligent Perceptionand Human-Machine Collaboration (ShanghaiTech University), the Shanghai Frontiers Science Center of Human-centered Artificial Intelligence. The experiments of this work were supported by the SIST computing Platform and HPC, ShanghaiTech University.

{
    \small
    \bibliographystyle{ieeenat_fullname}
    \bibliography{main}
}

\input{sec/X_suppl}

\end{document}

%% file: sec/0_abstract.tex
\begin{abstract}
Automated 3D scene generation is pivotal for applications spanning virtual reality, digital content creation, and Embodied AI. While computer graphics prioritizes aesthetic layouts, vision and robotics demand scenes that mirror real-world complexity which current data-driven methods struggle to achieve due to limited unstructured training data and insufficient spatial and physical modeling. We propose \ours{}, a diffusion-based framework that jointly learns spatial and physical relationships through a graph transformer, explicitly conditioning on posed scene point clouds for geometric awareness. Moreover, our model integrates differentiable guidance for collision avoidance, relational constraint, and gravity, ensuring physically coherent scenes without sacrificing relational context. Our experiments on 3D-FRONT and ProcTHOR datasets demonstrate state-of-the-art performance in spatial-relational reasoning and physical metrics. Moreover, \ours{} outperforms baselines in scene consistency and stability during pre- and post-physics simulation, proving its capability to generate simulation-ready environments for embodied AI agents.\footnotemark
\end{abstract}

\footnotetext{Our code and dataset are publicly available at \url{https://github.com/L-avenir/SPREAD}.}

%% file: sec/1_intro.tex
\section{Introduction}
\label{sec:intro}

\begin{figure*}[ht]
  \centering
  \includegraphics[width=0.99\textwidth]{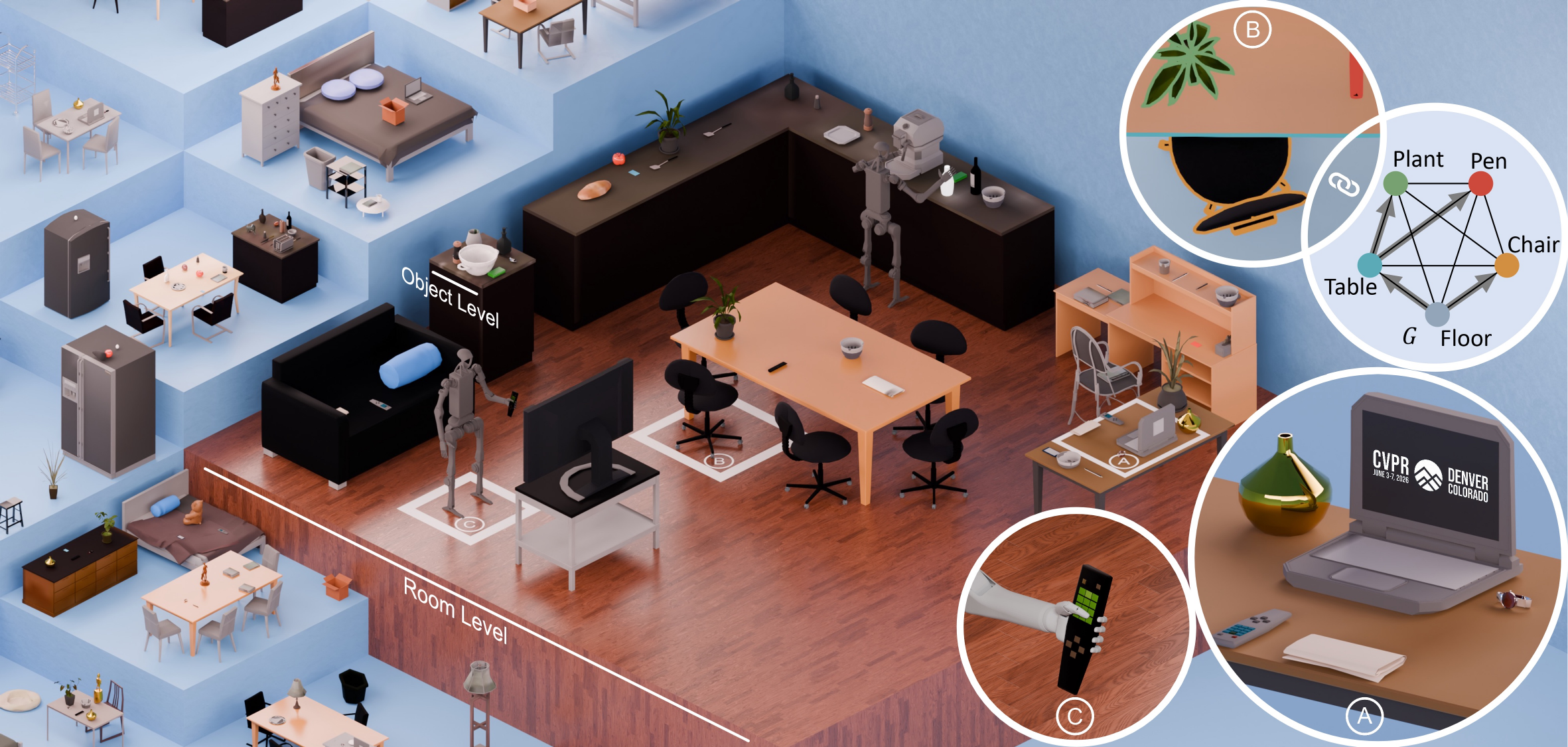}
  \label{fig:teaser}
  \caption{\textbf{Illustration of \ours{}}, a diffusion-based framework for generating physically plausible 3D scenes with rich object interactions. (A) \ours{} synthesizes detailed object-level layouts with natural spatial and physical interactions, going beyond coarse layout arrangements. (B) \ours{} faithfully adheres to provided spatial and physical graph priors, $\mathcal{G}$. (C) \ours{} can provide \textbf{simulation-ready} environments for embodied AI agents.}
\end{figure*}

Automated 3D scene generation is a critical task with applications ranging from virtual reality~\cite{ratnarajah2024listen2scene,li2022immersive, xu2023vr} and digital content creation~\cite{yao2025cast, hollein2023text2room, lin2023magic3d} to Embodied AI~\cite{li2024behavior, deitke2022, puig2023habitat}. Different domains place distinct requirements on the generated scenes. In computer graphics, the emphasis is often on geometry details, aesthetic object layouts and stylistic consistency, resulting in visually appealing and orderly environments. In contrast, 3D scenes used for training models in computer vision~\cite{wen2024foundationpose, fischer2024sama, thai20243} and robotics~\cite{gu2023maniskill2,zhang2024dexgraspnet} are expected to closely mirror the complexities of the real world. These scenes must accommodate cluttered arrangements, heavy occlusions, and diverse object poses to ensure the robustness of perception systems and embodied agents. Real-world environments can often appear "chaotic". For instance, a toy played by an infant may be placed in an arbitrary pose, without any discernible logic. Capturing such diversity and disorder remains a major challenge for current data-driven methods, primarily due to the lack of sufficiently varied and unstructured training data.

Humans possess strong spatial and physical reasoning abilities that allow them to interpret and navigate the often chaotic real world. For instance, people can infer the pose of an object based on its interactions and relationships with surrounding objects. A pencil, for example, is more likely to lie flat on the ground for stability, but it can also stand upright when supported by a holder. The same object can exhibit different stable poses depending on its physical context. Beyond these physical relationships, humans also leverage deeper layers of understanding—such as functionality (e.g., a cup must remain upright to hold water) and cultural conventions (e.g., the placement of a knife and spoon to signal satisfaction with a meal). To generate truly realistic 3D scenes, a generative model must learn to reason about this foundational layer of physics: how objects stably interact, support each other, and coexist spatially in a functionally and semantically coherent manner.

Previous methods based on optimization~\cite{zhao2021luminous, zhang2021mageadd, qi2018human} are capable of achieving physically plausible results for individual scenes but suffer from poor scalability. Procedural generation techniques employ handcrafted spatial rules, enabling efficient large-scale generation but introducing artificial biases that reduce real-world variability. Recent deep generative models learn scene distributions directly from data -- some methods~\cite{lin2024instructscene, tang2024diffuscene, yang2024scenecraft} incorporate spatial relations as graph priors from text prompts or user specifications, achieving controllability but often neglecting physical constraints (leading to floating objects or penetration artifacts). Alternative approaches~\cite{yang2024physcene} enforce physical plausibility through guidance mechanisms, but fail to maintain realistic relational context, resulting in physically stable yet layout-incoherent scenes. Furthermore, most methods rely on datasets like 3D-FRONT~\cite{fu20213d} that capture only coarse furniture arrangements, lacking the detailed object interaction data necessary for complex physical relationships.

To this end, we introduce \ours{}, a guided diffusion framework that takes a foundational step toward reliable 3D scene generation by learning to reason about both spatial and physical relationships. These relationships are represented as graphs and incorporated into the diffusion process via graph transformer. Unlike prior methods~\cite{yang2024physcene, tang2024diffuscene} that rely on implicit shape embeddings, our framework conditions on the posed scene point cloud at each diffusion step with a geometry-aware perceiver module. Furthermore, \ours{} systematically enforces fundamental physical principles—such as mesh-level collision avoidance, stable object support, and adherence to gravity—through a set of carefully designed differentiable guidance functions. This integrated design enables the generative process to satisfy key relational and physical constraints, effectively determining the placement and orientation of objects in a physically consistent manner.

Our experiments demonstrate that \ours{} achieves state-of-the-art performance in spatial reasoning and physical plausibility, notably exhibiting low mesh-level collision rates. The generated scenes are simulation-ready, requiring little to no post-processing to ensure physical stability. To highlight the model’s capability for fine-grained spatial modeling, we train and evaluate \ours{} on both 3D-Front and ProcTHOR~\cite{deitke2022}, the latter offering a rich diversity of small objects.

In summary, our main contributions are threefold:
\begin{itemize}
    \item We propose a novel diffusion model that jointly represents spatial and physical relationships as differentiable graph priors, enabling scene synthesis that is both semantically coherent and physically plausible.
    \item We formulate mesh-level collision avoidance, stable physical relations and gravity as differentiable guidance functions, ensuring generated scenes obey physical principles while preserving rich object interactions.
    \item \ours{} achieves SOTA performance on physical plausibility metrics. Moreover, it demonstrates strong spatial and physical stability after simulations, validated on both furniture-scale and fine-grained interaction datasets.
\end{itemize}

%% file: sec/2_related_work.tex
\section{Related work}
\label{sec:related}

\begin{figure*}[!htp]
  \centering
  \includegraphics[width=0.95\textwidth]{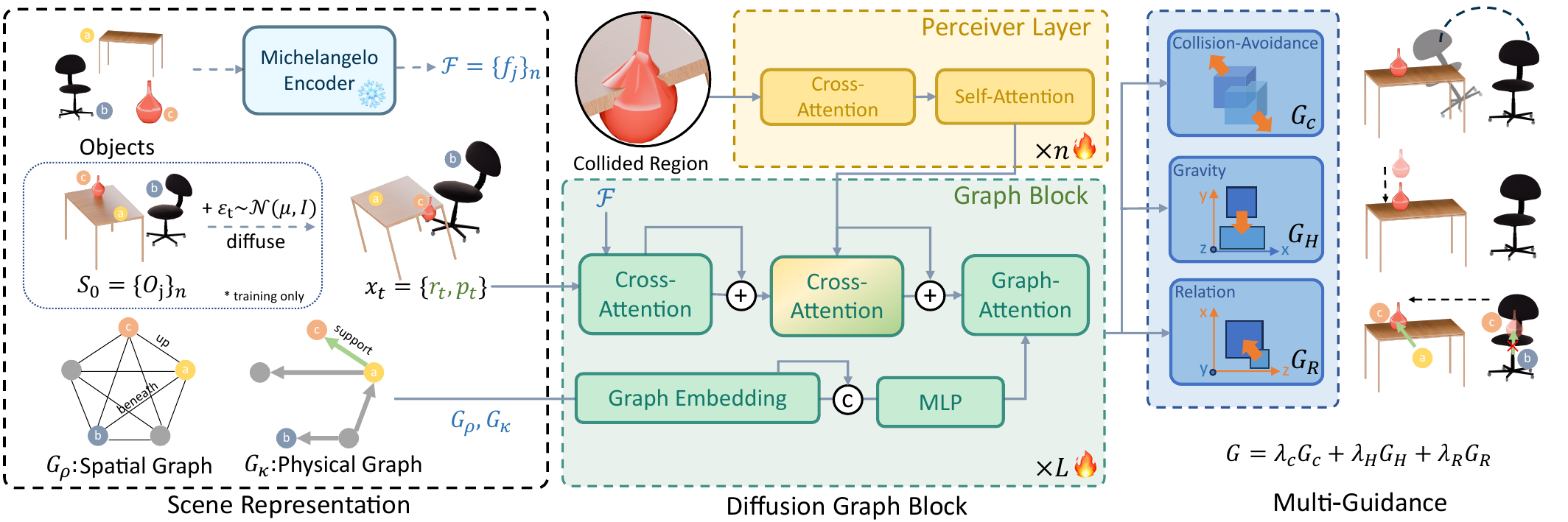}
  \caption{\textbf{Overview of \ours{}.} We propose \ours{}, a diffusion-based framework for generating physically plausible 3D scenes, which integrates relational constraints through spatial ($\mathcal{G_{\rho}}$) and physical graphs ($\mathcal{G_{\kappa}}$) while leveraging geometric perception via Perceiver Layers. The model employs graph-attention guided diffusion to jointly optimize physical plausibility and spatial relations during generation, producing realistic scenes with natural object interactions. }
  \label{fig:method}
\end{figure*}

\subsection{3D scene generation}
\paragraph{Procedural Scene Generation}
Procedural scene generation relies on predefined rules and is widely used in indoor design~\cite{raistrick2024infinigen}.
Prior work used statistical methods to derive object-distribution rules for structurally sound scene generation~\cite{chang2014learning, fisher2010context, chang2017sceneseer}. 
Other works cast generation as an optimization problem with constraints embedded in cost functions~\cite{zhang2021mageadd, deitke2022}.
Recently, large language models (LLMs) have introduced text-guided paradigms~\cite{feng2023layoutgpt, ling2025scenethesis, yang2024holodeck}.
These methods rely on manual rules, limiting their use in complex scenes. Our learning-based approach models rules from data, enabling more complex scene generation.

\paragraph{Graph-Driven Scene Generation}
Graph-driven methods represent scenes as object–relationship graphs, capturing interactions and guiding construction.
Semantic-spatial relation~\cite{wang2019planit}, dense relational~\cite{zhou2019scenegraphnet}, and hierarchical graphs~\cite{li2019grains} have been explored to capture spatial dependencies. 
Recent works further introduce commonsense-enhanced~\cite{zhai2023commonscenes} and language-guided scene graph~\cite{lin2024instructscene} to incorporate high-level semantics.
However, existing approaches primarily focus on modeling spatial relations, often neglecting the explicit incorporation of physical relations like "support". To address this limitation, our method enhances scene priors by introducing a physical relation graph, enabling more comprehensive and physically plausible scene generation.

\paragraph{Diffusion-Based Scene Generation}
Diffusion models are a leading generative paradigm, enabling high-quality and diverse synthesis via iterative noising-denoising~\cite{rombach2022high, ho2020denoising}.
In 3D scene generation, diffusion models enable object-level shape editing~\cite{zhai2023commonscenes} and occlusion-aware inpainting~\cite{wang2024perf}, showing promising results. 
Recently, they have been applied to full-scene synthesis, modeling entire 3D scene distributions for greater flexibility and expressiveness~\cite{tang2024diffuscene, bautista2022gaudi, yang2024physcene, lin2024instructscene, Hu2024MixedDF}. 
% However, existing methods rely on geometric features that lack pose reasoning, yielding scenes without hierarchy or realism.
% Our approach captures geometry and spatial relations under physical constraints, producing physically coherent scenes.
However, existing methods tend to focus on visual quality in 3D scenes, while paying less attention to both the consistency of object relationships and physical realism, leaving the joint modeling of these two aspects relatively unexplored in diffusion-based approaches.

\subsection{Guided diffusion} 
Guided diffusion models use external signals to steer generation, enabling fine-grained control and alignment with objectives~\cite{nichol2021glide, dhariwal2021diffusion, ho2022classifier, pfaff2025steerable}. 
Recent 3D scene generation studies have adopted guided diffusion models. 
SceneDiffuser~\cite{jiang2024scenediffuser} conditions on physical constraints, while Physcene~\cite{yang2024physcene} integrates physical and interaction cues.
Unlike Physcene, which relies on bounding-box representations, our framework leverages mesh- and relation-level guidance to enable finer-grained, physically consistent control and to produce more realistic, structurally complex scenes.

\subsection{Physical constraints}
Physical constraints are physics-based rules (e.g. motion laws and object interactions) that ensure plausibility in simulation.
Simulation-based 3D modeling uses physics engines to enforce hard constraints for realism~\cite{todorov2012mujoco, makoviychuk2021isaac, xiang2020sapien}.
Motion capture uses physics-informed losses to enhance realism, plausibility, and temporal coherence~\cite{xie2021physics, huang2022neural, ju2023physics}.
In generative models, physical constraints are used as conditional signals~\cite{gillman2025force} or latent embeddings~\cite{wu2024thor} to enforce physical consistency.
Our method guides scene generation with physical constraints, ensuring consistency and coherence for realistic, complex 3D scenes.

%% file: sec/3_method.tex
\section{Method}
\label{sec:method}

To generate physically plausible scenes, we propose \ours{}, an integrated framework that combines our proposed geometric perceiver layers with guided diffusion. In section~\ref{subsection:scene_representation}, we outline the compositional elements of our scene representation, especially geometric and relational priors. In section~\ref{subsection:diffusion_for_scene_modeling} and section~\ref{subsection:model_architecture}, we detail our model architecture which integrates Perceiver Layer - a dedicated geometric perception module that enables the network to learn geometric constraints during training. In section~\ref{subsection:multi_guidance}, for posterior optimization during inference, we propose a novel combination of diffusion guidance mechanism that simultaneously addresses physical plausibility and relational constraints.

\subsection{Scene representation}
\label{subsection:scene_representation}

To enable comprehensive modeling and generation of physically plausible scenes, our framework relies on structured representations that precisely capture objects and their spatial interactions as priors.

As illustrated in Fig~\ref{fig:method}, the scene $S_i$ contains objects $o_j^i$ each defined by a tuple $\langle p_j^i, r_j^i, f_j^i, \rho_j^i, \kappa_j^i \rangle$, where 3D translation $p_j^i \in \mathbb{R}^3$ stands for the centroid position of $o_j^i$,  orientation $r_j^i \in \mathtt{SO}(3)$. Moreover, the geometric features $f_j^i \in \mathbb{R}^d$ provide a $d$-dimensional shape descriptor. The spatial relations $\rho_j^i$ model pairwise relative directions. For example, $\rho(o_k^i, o_l^i)$ indicates if $o_l^i$ is left, right, front, or back of $o_k^i$. Similarly, the physical interactions $\kappa_j^i(o^i_k, o_l^i)$ describe support, contact, or attachment relation between objects, which enables the generation with more comprehensive modeling and additional controllability. Notably, instead of using images as visual references ~\cite{li2024genrc,chatterjee20253d}, we employ explicitly structured graph representations to separately model spatial relationships $\mathcal{G}_{\rho} \in \{0, ..., m\}^{N\times N}$ and physical interactions $\mathcal{G}_{\kappa} \in \{0, 1, ..., q\}^{N\times N}$ as latent constraints, where $m$ and $q$ represent the number of spatial relations and physical interactions respectively.Technically, we utilize a continuous representation~\cite{hempel20226d} to parameterize rotation $r$, where $r \in \mathbb{R}^{6}$.

\subsection{Geometry-aware diffusion modeling}
\label{subsection:diffusion_for_scene_modeling}
Here, we introduce \ours{}, our graph-based diffusion for scene generation with geometric awareness. As illustrated in Fig~\ref{fig:method}, \ours{} differs from existing scene generation methods by explicitly modeling physical and spatial relations between denoised meshes at each denoising step, while incorporating additional geometric inputs through a geometry-aware perceiver module. Furthermore, we employ a graph transformer with cross-attention blocks to parameterize $\epsilon_{\theta}(\mathbf{x}_t, t, \mathbf{f}, \mathcal{G}_\rho, \mathcal{G}_\kappa)$, where $t$ denotes time embeddings, $\mathbf{f}$ denotes geometric features for collections of objects in the scene, $\mathcal{G}_\rho$ and $\mathcal{G}_\kappa$ represent spatial and interaction relation graph, respectively.

The diffusion process operates on a structured scene representation $\mathcal{S}_i = \{o_j^i\}_{j=1}^N$, where each object is parameterized as $\langle p_j^i, r_j^i, f_j^i, \rho_j^i, \kappa_j^i \rangle$. We construct a joint state space by concatenating scene representations across all objects,
\begin{equation}
    % \scriptsize
    \mathbf{x}_0 = \bigoplus_{j=1}^N [ p_j^i \Vert r_j^i ] \in \mathbb{R}^{N \times (3+6)} 
\end{equation}
forming the basis for the diffusion process. The forward process follows a Markov chain that gradually perturbs the data through Gaussian transitions, preserving the topological structure while adding noise 
\begin{equation}
% \scriptsize
    q(\mathbf{x}_t|\mathbf{x}_{t-1}) = \mathcal{N}(\mathbf{x}_t; \sqrt{1-\beta_t}\mathbf{x}_{t-1}, \beta_t\mathbf{I})
\end{equation} where $\beta_t$ stands for the variance of the Gaussian noise added at each step of the forward (diffusion) process.

For reversal process, we model the spatial relations $\mathcal{G}\rho$ and physical relations $\mathcal{G}\kappa$ between objects as graph structures, where each graph is represented by an adjacency matrix of shape $(N, N)$, with elements indicating K possible relation types. These graph structures are first mapped to a continuous latent space through embedding layers,
\begin{equation}
    \mathbf{E} = \text{MLP}(\text{Embedding}(\mathcal{G}))
\end{equation} where edge embeddings $\mathbf{E} \in \mathbb{R}^{N\times N \times d_e}$. These embeddings are then injected as bias terms into the graph attention layers. At each denoising step t, the graph structures condition the generation process through the diffusion graph block 
\begin{equation}
    \mathbf{H}{t}^{l+1} = \text{GraphBlock}^l(\mathbf{H}t^l, \mathcal{G}\rho, \mathcal{G}\kappa)
\end{equation} while simultaneously processing both types of relational information.

By enabling relational modeling in continuous feature space through MLP projection, \ours{} achieves joint optimization of spatial relations and physical constraints in each graph block layer.

\begin{table*}
\caption{\textbf{Quantitative comparison.}
Our method matches baseline FID on 3D-FRONT while setting new state-of-the-art on ProcTHOR: it dramatically reduces mesh collisions, achieves the highest graph recall (GRecall), minimizes average support distance (ASD), and delivers the greatest scene stability under Isaac Sim.}
  \centering
  \footnotesize
  \begin{tabular}{l
                  *{3}{c}
                  c
                  *{5}{c}
                  c}
    \toprule
    \multirow{4}{*}{Method}
      & \multicolumn{4}{c}{3D-FRONT}
      & \multicolumn{5}{c}{ProcTHOR} \\
    \cmidrule(lr){2-5} \cmidrule(lr){6-10}
    & \multicolumn{3}{c}{$\text{Col}_{\text{mesh}}\downarrow$}
      & \multirow{3}{*}{FID$\downarrow$}
      & \multirow{3}{*}{GRecall$\uparrow$}
      & \multirow{3}{*}{$\text{Col}_{\text{mesh}}\downarrow$}
      & \multirow{3}{*}{ASD$\downarrow$}
      & \multirow{3}{*}{Stability$\uparrow$}
      & \multirow{3}{*}{FID$\downarrow$}  \\ 
    \cmidrule(lr){2-4} 
    & Bedroom & Livingroom & Diningroom
      & 
      &  &  &  &  & 
      &  \\ 
    \midrule
    ATISS
      & 0.275 & 0.451 & 0.428
      & 68.0
      & / & 0.174 & 0.510 & 0.813 & 33.9 \\
    DiffuScene
      & 0.298 & 0.359 & 0.376
      & 61.6
      & / & 0.360 & 0.071 & 0.886 & 21.4 \\
    InstructScene
      & 0.285 & 0.350 & 0.331
      & \textbf{61.3}
      & 0.964 & 0.260 & 0.021 & 0.876 & 20.0 \\
    Ours
      & \textbf{0.097} & \textbf{0.185} & \textbf{0.183}
      & 64.7
      & \textbf{0.979} & \textbf{0.121} & \textbf{0.007} & \textbf{0.950} & \textbf{18.8} \\
    \bottomrule
  \end{tabular}
  
  \label{tab:merged_metricD}
\end{table*}

\begin{figure*}[!t]
  \centering
  \includegraphics[width=\textwidth]{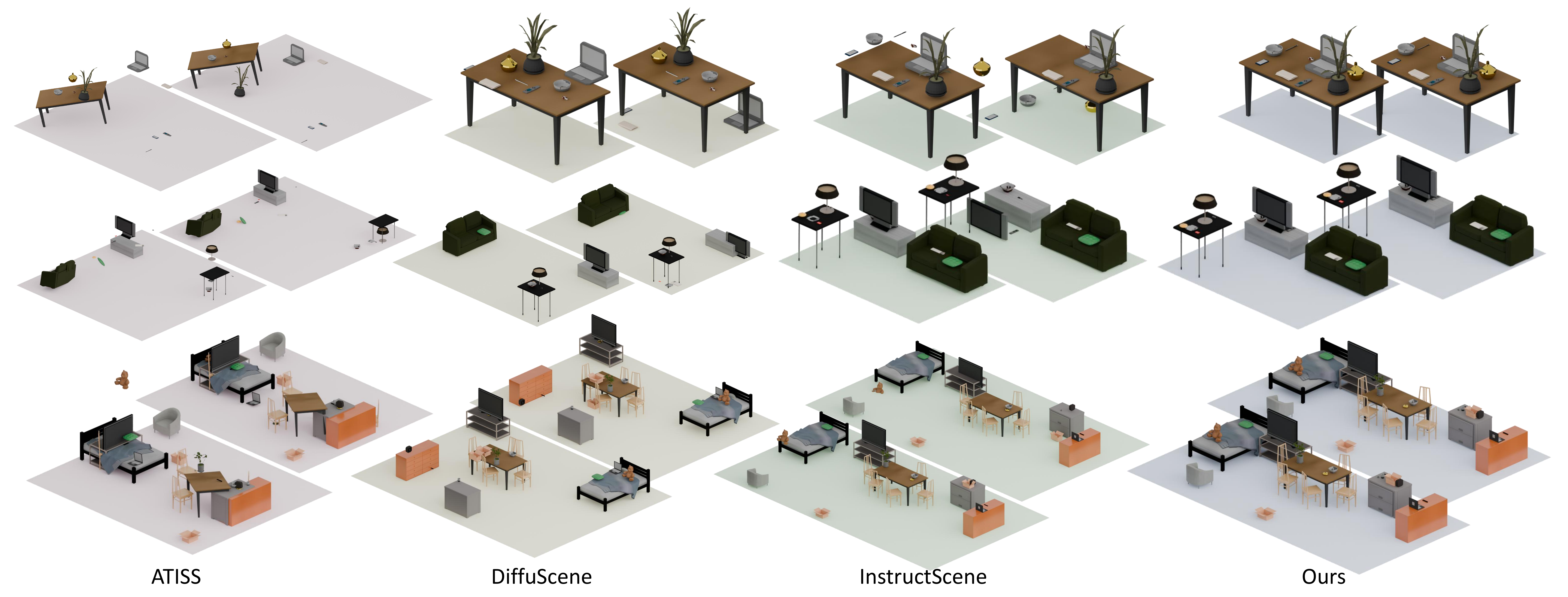}
  \caption{\textbf{Comparative Generation and Simulation Results.} Visual comparison of scene layouts produced by our method versus three baseline approaches, shown before (left) and after (right) physics simulation.}
  \label{fig:comparison}
\end{figure*}

\subsection{Model architecture}
\label{subsection:model_architecture}
% @kuixiang
Our geometry-aware diffusion network jointly models object orientation, geometry, and inter-object relations as multimodal priors to transform a random layout into a physically plausible, semantically coherent 3D scene. As described in section~\ref{subsection:diffusion_for_scene_modeling}, each object node is represented by a 9-dimensional state vector $\mathbf{x}_i$. This vector is then augmented with fixed sinusoidal positional encodings and projected into the attention dimension through a linear layer. Concurrently, each object is encoded by a pretrained Michelangelo encoder~\cite{zhao2023michelangelo} into 256 64-dimensional tokens, which are then linearly projected to the attention dimension. These shape tokens remain constant during diffusion, providing a stable shape prior.

At each diffusion timestep \(t\), dynamic geometric interactions are captured by sampling \(M\) points \(\mathbf{p}_{i}^{M}\) on the noisy mesh of object \(i\) and computing the one-way Chamfer distance~\cite{fan2017point} to all other objects’ point clouds $\mathcal{P}_{\neg i}$. We assign a sign to each distance via the nearest neighbor’s normal \(\mathbf{n}_{\mathrm{nn}}\), thereby approximating a signed distance field:
\begin{equation}
% \scriptsize
d_{\mathrm{scd}}(\mathbf{p})
=
\min_{\mathbf{q}\in\mathcal{P}_{\neg i}}\|\mathbf{p}-\mathbf{q}\|_2 \cdot
\mathrm{sign}\bigl(\mathbf{n}_{\mathrm{nn}}^\top(\mathbf{p}-\mathbf{q})\bigr)
\end{equation}
This procedure yields a feature tensor of shape \((B,N,M,4)\), where the first three channels encode global coordinates and the fourth channel encodes \(d_{\mathrm{scd}}\). A Perceiver~\cite{jaegle2021perceiver} module then distills these sparse, high-dimensional features into n \(d\)-dimensional tokens \(\mathbf{f}^{\mathrm{geo}}\) via cross-attention, enabling the network to perceive collisions and penetrations.

Discrete spatial and physical relations (e.g., “left of,” “supports”) are embedded and concatenated as edge features \(\mathbf{e}_{\rho\kappa}\), together with node features forming an explicit scene graph. We stack \(L\) multimodal graph layers to iteratively fuse and propagate information: within each block, node features first attend to static shape tokens and then to dynamic geometric embeddings via sequential cross-attention, producing shape- and geometry-aware fused representations; these representations and the edge features are subsequently processed by a graph attention block, in which multi-head graph attention propagates signals along explicit edges. All normalization layers condition on the timestep embedding $\mathbf{t}_{\mathrm{emb}}$ via AdaLayerNorm~\cite{peebles2023scalable, perez2018film}, making the denoising process time-aware. Finally, the refined node features are projected by a MLP to predict the noise \(\hat\epsilon\), and training minimizes the mean squared error \(\|\hat\epsilon - \epsilon\|^2\). By deeply integrating static shape priors, dynamic Chamfer-based geometry, and explicit relational structure, our framework achieves high-fidelity, physically consistent 3D scene generation.

\subsection{Multi-guidance framework}
\label{subsection:multi_guidance}

The proposed diffusion guidance framework incorporates physically-grounded constraints through differentiable operators that modify the score function of the diffusion process. Formally, given a diffusion model with learned score function $s_\theta(\mathbf{x}_t,t)$, the guided reverse process follows:

\begin{equation}
\nabla_{\mathbf{x_t}} \log p_\gamma(\mathbf{x_t}) = s_\theta(\mathbf{x_t},t) + \gamma \nabla_{\mathbf{x_t}} \mathcal{G}(\mathbf{x_t})
\end{equation}
where $\mathcal{G}(\mathbf{x_t})$ represents our composite guidance signal combining three key components:
collision guidance, gravity guidance, and relation guidance. 
With the weight of each guidance, the overall guidance is defined via:
\begin{equation}
% \scriptsize
    \mathcal{G} = \lambda_{C} \mathcal{G}_{C} + \lambda_{H} \mathcal{G}_{H} + \lambda_{R} \mathcal{G}_{R}
\end{equation}

Together, these guidance terms enforce consistency with real-world physical laws and foster the generation of stable, contact-rich 3D scenes.

\begin{figure}[t!]
  \centering
  \includegraphics[width=\linewidth]{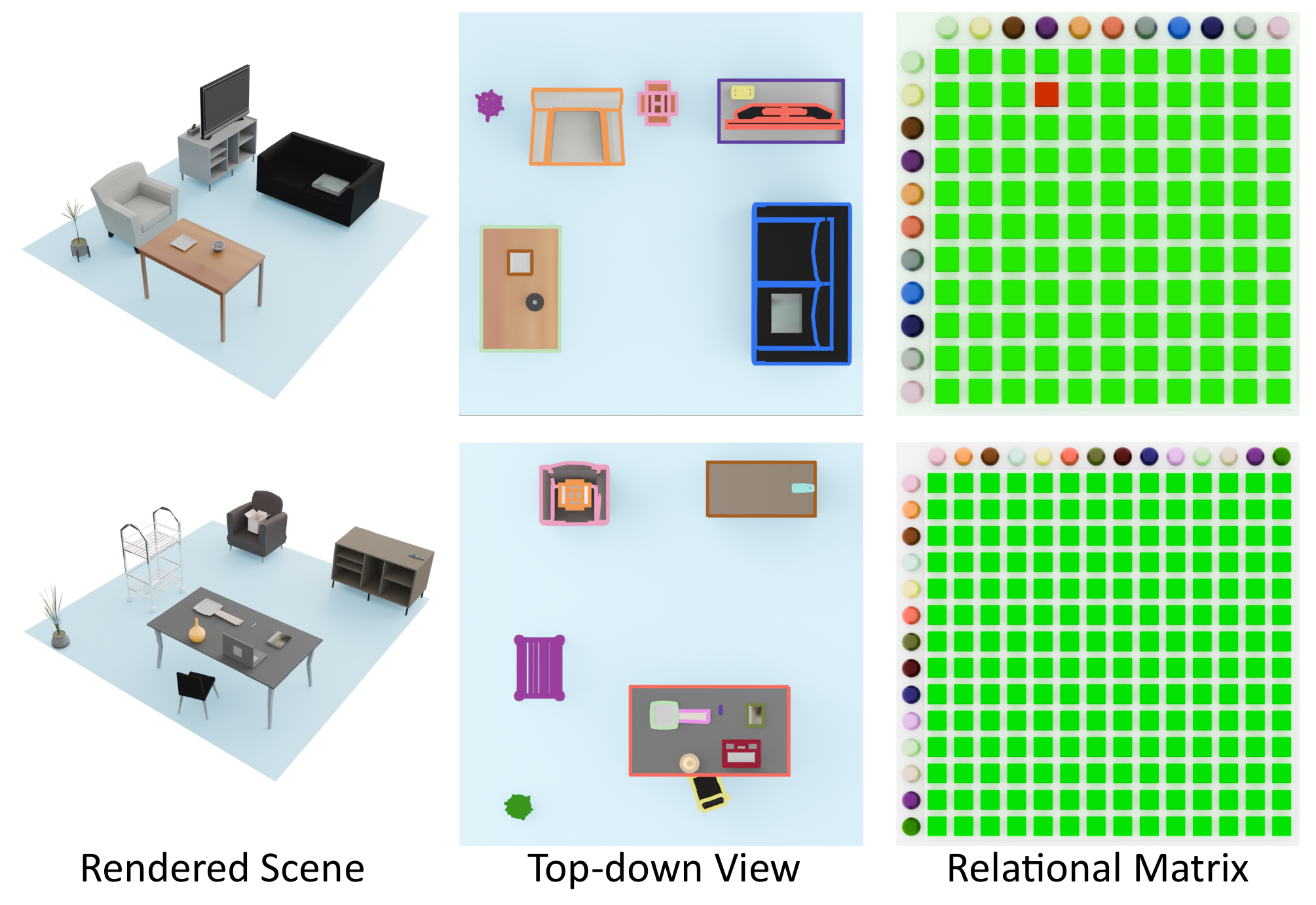}
  \caption{\textbf{Scene \& Relation Visualization.} For two generated scenes, we show the final render (left), the top-down layout (middle), and the pairwise relation evaluation matrix (right). The matrix encodes every object-pair’s spatial relation: green entries denote correct relations (w.r.t. the ground truth), and red entries denote incorrect ones.}
  \label{fig:relation_demo}
\end{figure}

\paragraph{Collision Guidance}
We introduce a collision-avoidance objective function based on intersecting triangles from different meshes. Unlike approaches like Physcene~\cite{yang2024physcene} which take predicted bounding boxes as  approximation, our method directly quantifies collision relationships based on mesh triangles, which proves more efficient and accurate. Specifically, we use $a$ and $b$ to represent any two distinguish objects in the scene. And we use $t^i_a$ and $t^j_b$ to denote the $i$-th triangle of $a$ and the $j$-th triangle of $b$. We use $\mathbf{CoDF}$~\cite{tzionas2016capturing} to evaluate the mesh-based collision-free guidance via:  

\begin{equation}
% \scriptsize
    \mathcal{G}_{C}=\frac{1}{|C|}\sum_{a,b,a\neq b}\sum_{(i,j)\in C}\mathbf{CoDF}(t_a^i, t_b^j)
\end{equation}
where $C$ represents collision triangle pairs found with BVH, and $\mathbf{CoDF}$ represents the conical distance field. The overall penalty is obtained by summing the penalties with all collision pairs of surface patches from different objects.

\paragraph{Gravity Guidance}

To ensure physically plausible support relationships, we model gravity constraints by computing vertical distances between objects and their supporters. For each object, we compute the vertical offset $r_i = d_i - \epsilon$, where $d_i$ is the distance from the object to its supporter, and $\epsilon$ is an empirical minimal threshold preventing objects from potential intersection caused by gravity guidance. This formulation naturally penalizes both excessive floating ($r_i > \theta_H$) and interpenetration ($r_i < 0$) while allowing small deviations within tolerance $\theta_H$.
\begin{equation}
% \scriptsize
    \mathcal{G}_{H}=\sum_{r_i>\theta_H\ \vee \ r_i<0}|r_i|\\
\end{equation}

\paragraph{Relation Guidance}

In reality, objects exhibit complex interrelations. To model these, we introduce a score function based on the extent of overlap between their projections onto the XZ‑plane. This approach is both effective and efficient, since we approximate each object by its projection convex hull. Instead of calculating the exact overlap area, we estimate a penalty by measuring the distances from all the object vertices lying outside their supporting object to the supporting hull. To be more specific, we denote a directed pair $(i, j)$ to indicate that the object $i$ is supported by the object $j$, and we use $E$ to represent the set of all such relations. Additionally, we denote $V_{i,j}$ to represent all the vertices of the $i$-th object's projection outside the convex hull of the $j$-th object. Thus, we further define the guidance as follows:
\begin{equation}
    \mathcal{G}_R=\sum_{(i,j)\in E}\sum_{\alpha \in V_{i,j}}\frac{\mathbf{s}(\alpha, j)}{|V_{i,j}||E|}
\end{equation}
where $\mathbf{s}(\alpha, j)$ represents the minimal Euclidean distance between a point $\alpha$ to the convex hull of supporter $j$.

%% file: sec/4_experiments.tex
\begin{table}[!t]
  \caption{\textbf{Ablation on ProcTHOR.} starting from our base model, adding the geometry module and then the full multi-guidance framework yields consistent improvements in all physical metrics.}
  \centering
  \footnotesize
  \setlength{\tabcolsep}{4pt}
  \begin{tabular}{lccccc}
    \toprule
    Method         & GRecall$\uparrow$ & $\text{Col}_{\text{mesh}}\downarrow$ & ASD$\downarrow$ & Stability$\uparrow$ \\
    \midrule
    Ours      & 0.963     & 0.241   & 0.014 & 0.934      \\
    +Geometry    & 0.965     & 0.225   & 0.012 & 0.938      \\
    +Guidance    & \textbf{0.979}     & \textbf{0.121}   & \textbf{0.007} & \textbf{0.950}      \\
    \bottomrule
  \end{tabular}

  \label{tab:ablation_procthor}
\end{table}

\section{Experiments}
\label{section:exp}

\paragraph{Datasets}
We evaluate our model on two large-scale indoor datasets: (1) 3D-FRONT~\cite{fu20213d} and (2) ProcTHOR~\cite{deitke2022}, which together capture both aesthetic layouts and rich physical interactions. 3D-FRONT provides high-quality, designer-curated scenes; we adopt the InstructScene~\cite{lin2024instructscene} preprocessing pipeline, augmenting each scene with explicit relative-position annotations (e.g., "left-of", "above") to form structured scene graphs. To better model complex object interactions, we utilize ProcTHOR's procedurally generated indoor scenes, excluding non-supportive meshes (e.g., wall hangings). We then apply physics-based corrections to resolve interpenetration and remove floating objects, while annotating spatial relationships using the same scheme as in 3D-FRONT. 

\begin{figure*}[!t]
  \centering
  \includegraphics[width=0.97\textwidth]{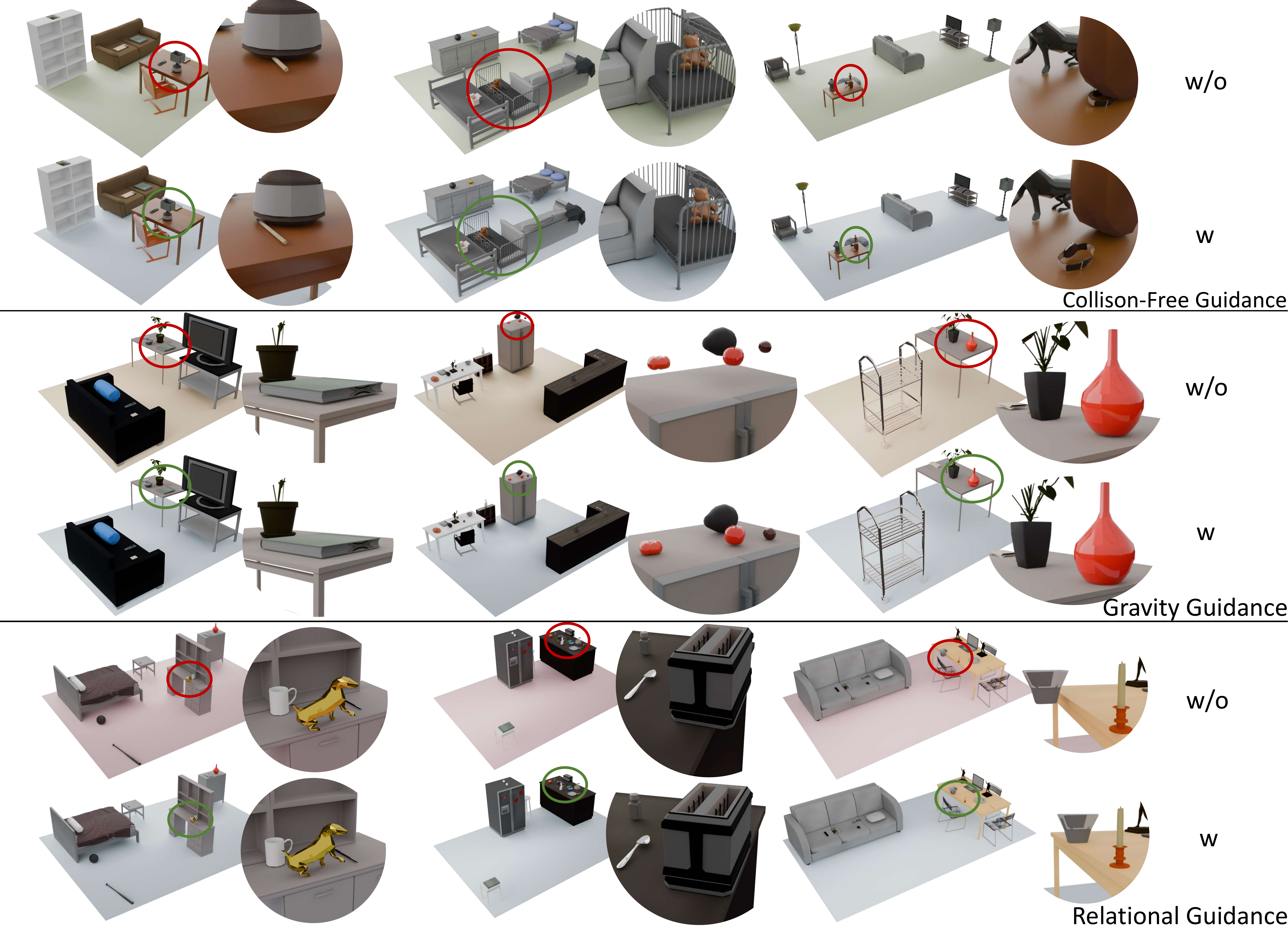}
 \caption{\textbf{Guidance Ablation.} Results showing effect of different guidance terms. Each major row compares results before (top) and after (bottom) adding a specific guidance. Columns show different scenes. Red circles highlight issues such as collisions, floating, or incorrect spatial relations before guidance; green circles show improvements after applying guidance, with zoom-in views for clarity.}
  \label{fig:guidance_ablation}
\end{figure*}

\paragraph{Baselines}
We compare \ours{} against three baselines: (1) ATISS~\cite{paschalidou2021atiss}, a permutation‐invariant transformer that  models scene generation as an autoregressive process over an unordered set of objects. (2) DiffuScene~\cite{tang2024diffuscene}, a model that adopts a denoising diffusion probabilistic model to generate scenes in a non-autoregressive manner. (3) InstructScene, which introduces a two-stage, graph-based framework designed for instruction-driven synthesis. These baselines provide a robust benchmark covering autoregressive, diffusion-based, and graph-structured generative methodologies. We refer the reader to the Supplementary Material for more information.

\paragraph{Metrics}
To ensure a comprehensive evaluation of our model, we assess visual quality, physical plausibility, and structural fidelity. For visual fidelity and diversity, we use the \emph{Fréchet Inception Distance} (FID). Physical plausibility is measured via the \emph{Mesh Collision Rate}, which quantifies object intersections. For the ProcTHOR dataset, we introduce three specialized metrics: \emph{Graph Recall} (GRecall) to evaluate structural accuracy by comparing inferred spatial relationships, \emph{Average Support Distance} (ASD) to assess contact surface quality through signed distance functions, and \emph{Stability}, which is measured by simulating scenes in NVIDIA Isaac Sim~\cite{makoviychuk2021isaac} and verifying object relationship consistency. Metric details are attached in the supplementary material.

\subsection{Results on scene generation}

Our framework’s advantages become especially clear when evaluated on the interaction-rich ProcTHOR dataset. As Table~\ref{tab:merged_metricD} shows, although we match baseline FID on 3D-FRONT, we surpass all existing approaches on every physically grounded metric in ProcTHOR. In particular, our method dramatically reduces $\text{Col}_{\text{mesh}}$ and achieves a GRecall of 0.979—demonstrating faithful adherence to the true scene layout (Figure~\ref{fig:relation_demo}). It yields an Average Support Distance of just 0.007, indicating virtually gap-free contact.

\subsection{Ablation study}

Our ablation study on the ProcTHOR dataset evaluates the separate contributions of the geometry-aware perceiver module and multi-guidance framework (Table~\ref{tab:ablation_procthor}). The baseline vanilla diffusion model shows substantial improvement when augmented with the geometry-aware perceiver (+Geometry), evidenced by reduced collision rate and ASD metrics. The complete multi-guidance framework (+Guidance) combined with geometry-aware diffusion achieves the most significant gains, reaching state-of-the-art physical plausibility. As Figure~\ref{fig:guidance_ablation} illustrates: collision guidance effectively removes interpenetrations, gravity guidance eliminates floating artifacts, and relational guidance maintains proper support structures - collectively achieving the lowest ASD, highest GRecall, and best simulation stability.
\subsection{Scene consistency in pre-post simulation} 

To rigorously evaluate physical plausibility, we measure scene stability under NVIDIA Isaac Sim. As Table~\ref{tab:merged_metricD} shows, our method achieves an Isaac Stability score of 0.950, meaning the vast majority of pairwise object relationships remain intact after simulation. Figure~\ref{fig:comparison} offers a side-by-side comparison: while baseline layouts frequently suffer object displacement and structural drift under physical forces, ours remain virtually unchanged, underscoring the practical benefit of integrating physical reasoning directly into the generation process.
\subsection{User study} 

\begin{figure}[th!]
  \centering
  \includegraphics[width=0.94\linewidth]{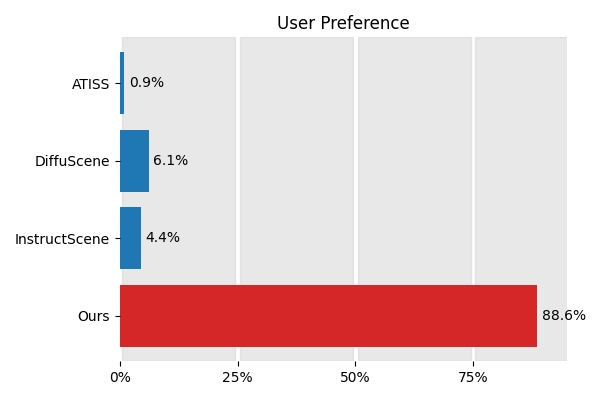}
  \caption{\textbf{User Study.} Our method dominated with 88.6\% of the votes, above ATISS with 0.9\%, InstructScene with 4.4\%, and DiffuScene with 6.1\%, indicating that our method better preserves physical consistency and scene rationality.}
  \label{fig:userstudy}
\end{figure}

We have conducted a user study focusing on physical consistency, assessing adherence to physical laws, and scene rationality, evaluating high-level semantic and commonsense coherence. We randomly selected five scenes, each containing a minimum of 10 and a maximum of 22 objects. A total of 57 valid responses were collected, each evaluating randomly selected three scenes to determine the preferred method. As illustrated in Figure~\ref{fig:userstudy}, our method garnered 88.6\% of votes, thus underscoring its superiority over ATISS(0.9\%), DiffuScene(6.1\%) and InstructScene(4.4\%). This finding suggests that our method generates scenes that are more physically consistent, scene-rational, and show greater consistency with human preferences.

\subsection{Inference speed comparison}

In this section, we analyze the computational efficiency of state-of-the-art methods. The average inference latency per scene is as follows: ATISS (0.02s), InstructScene (2.58s), Diffuscene (10.25s), and SPREAD (14.72s). The higher latency of our method is an inherent trade-off of its generative architecture, deliberately designed to model complex object relationships and generate coherent scenes, prioritizing quality and structural integrity over inference speed.

\begin{table}[!t]
  \caption{\textbf{Comparison of inference times }(in seconds) across different scene generation methods.}
  \centering
  \footnotesize
  \setlength{\tabcolsep}{4pt}
  \begin{tabular}{lccccc}
    \toprule
    Method         & \ours{} & ATISS & InstructScene & DiffuScene \\
    \midrule
    Inference Time (s)      & 14.72     & 0.02   & 2.58 & 10.25 \\
    \bottomrule
  \end{tabular}

  \label{tab:inference_latency}
\end{table}

%% file: sec/5_conclusion.tex
\section{Conclusion} \label{section:conclusion}

We present \ours{}, a guided diffusion framework that jointly models object spatial and physical relationships through differentiable graph priors and multi-guidance mechanism. Our method synthesizes  physically plausible and simulation-ready scenes. Experiments demonstrate superior performance in spatial reasoning and physical metrics, with robustness validated under simulations. 
\vspace{-10pt}
\paragraph{Future work} 
While effective at preserving physical constraints, our method is currently limited to indoor scenes due to dataset availability. Future work will extend to outdoor generation by leveraging image-conditioned paradigms~\cite{yao2025cast, ling2025scenethesis}. To address the computational cost inherent in our iterative diffusion process, we will explore efficient alternatives like flow matching. Furthermore, we will investigate formulating the diffusion directly on the SE(3) manifold to better leverage its geometric prior for more principled scene synthesis.

%% file: sec/X_suppl.tex
\clearpage
\setcounter{page}{1}
\maketitlesupplementary

\section{Implementation details}
\subsection{Hyper parameters}
All experiments were conducted on NVIDIA H20 GPUs. The model was trained for $2,000$ epochs with a batch size of $16$, taking approximately $50$ GPU hours. We employ the AdamW optimize with a learning rate of $1 \times 10^{-4}$ and a weight decay of $1 \times 10^{-4}$. Our diffusion backbone adopts the Denoising Diffusion Probabilistic Models (DDPM) framework. We utilize \texttt{squaredcos\_cap\_v2} noise schedule, operating over a total of 1,000 denoising timesteps. As mentioned in Section~\ref{sec:method}, we sample $M$ points from mesh faces with noisy rotation and translation to compute the signed chamfer distance, which yields feature vectors during training. This strategy significantly enhances the model's capacity to capture geometric details and contributes to spatial-physically reasoning and scene generation. Here we empirically set M to be 2,000. During sampling, we integrate three distinct guidance terms to steer the generation towards physically plausible and relationally consistent scenes. We empirically set our guidance weights    $\lambda_C$,$\lambda_R$, $\lambda_H$ for collision, relation, gravity guidance to be $7.5 \times 10^{-3}$, $1.0 \times 10^{-3}$, $1.0 \times 10^{-3}$, respectively.

During the quantitative evaluation, we compute the Fréchet Inception Distance (FID) on CLIP features using the Clean-FID\footnote{https://pypi.org/project/clean-fid/} library. We employ Nvidia Isaac Sim with version 5.1.0 to evaluate simulation stability.

\subsection{Shape Encoder}
We leverage the pre-trained shape encoder, \textbf{Michelangelo}~\cite{zhao2023michelangelo}, to extract latent shape codes for CAD models from 3D-FRONT~\cite{fu20213d} dataset and the processed ProcTHOR~\cite{deitke2022} dataset. It functions as an image-text-aligned 3D shape Variational Auto-Encoder. Here we briefly introduce the forward process of the encoder. First, we initially sample a point cloud $P \in \mathbb{R}^{N \times 3}$ from the surface of the 3D shape. Then the points are fed into the \textbf{SITA-VAE} module of \textbf{Michelangelo}, which captures both low-level geometric information via $L$ query tokens and high-level semantic information via a single global head token. Finally, multiple self-attention layers iteratively refine the representation and obtain the final embeddings $f \in \mathbb{R}^{N \times L \times D}$. The embeddings are used directly for the following diffusion training, serving as a compressed and semantically rich geometry representation of shapes.

\subsection{Geometry-aware module in diffusion}
In the main paper, we introduced a novel training strategy that endows diffusion models with explicit awareness of geometry and inter-relationship between paired objects in the scene. During the diffusion training process, we aim at enabling the model to perceive collisions and penetrations in the scene. To achieve this, we sample $M$=2,000 points from the surface for each noisy-posed object and compute the one-way Chamfer distance to all other objects’ point clouds. This procedure divides the points into two subsets: the collided points and the remains. Rather than relying on a binary collision indicator \{0, 1\}, we preserve the signed distance value as a continuous feature. Consequently, a feature tensor of shape $(N, M, 4)$ is obtained, which represents the collision information (referred to as relational information) for the scene at timestep $t$. Then a lightweight yet effective perceiver-based transformer is employed to convey the explicit geometric information of the scene at timestep $t$.

\section{Dataset}
\subsection{The insufficiency of public dataset}
\begin{figure*}[htbp]
  \centering
        \begin{minipage}[b]{0.495\textwidth}
              \includegraphics[width=\textwidth]{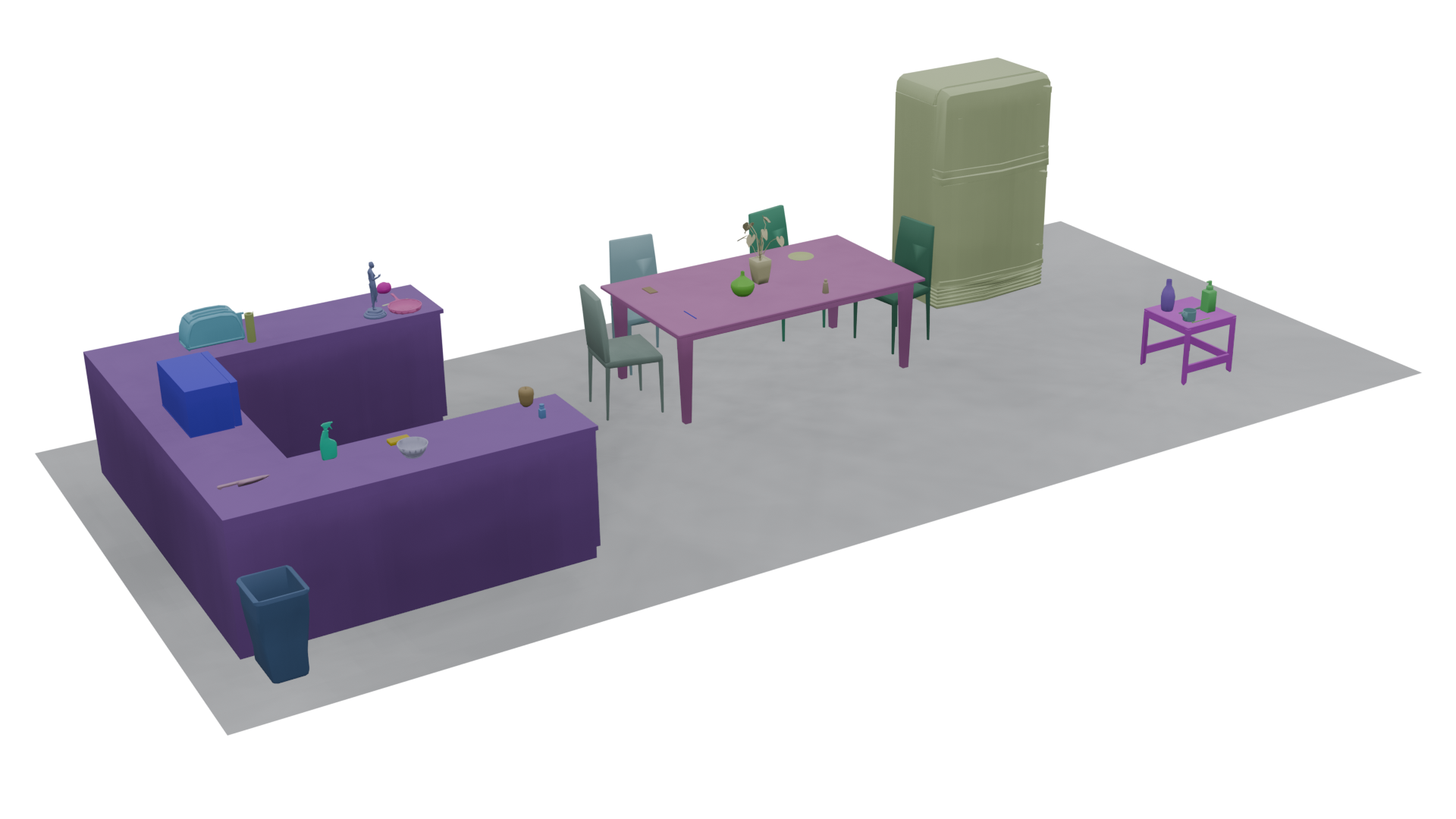}
        \end{minipage}
        \hfill
        \begin{minipage}[b]{0.495\textwidth}
            \includegraphics[width=\textwidth]{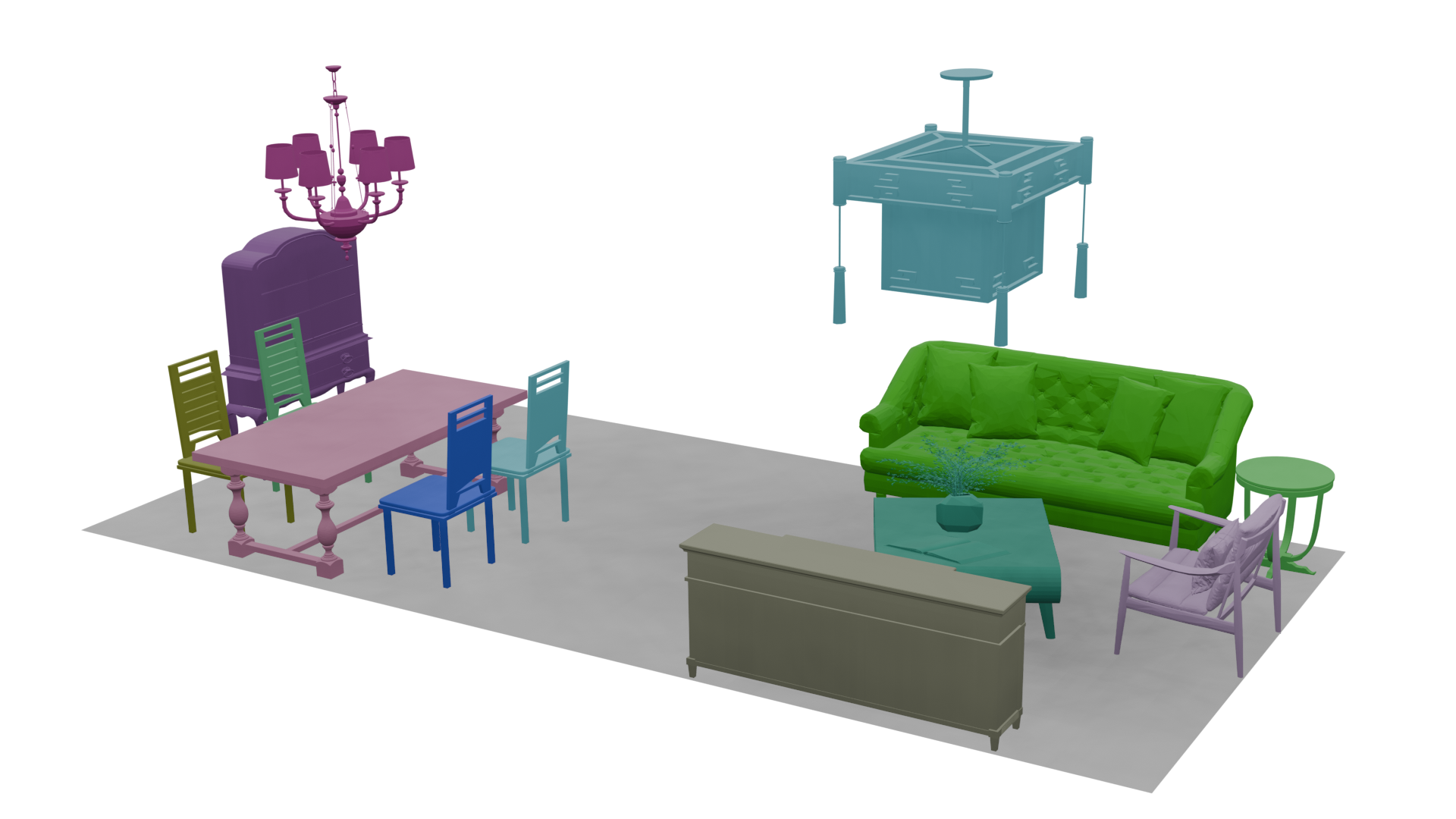}
        \end{minipage}
        \caption{\textbf{Comparison of relational complexity between our pre-processed dataset (left) and the original 3D-FRONT dataset (right).} The sample from our dataset (left) contains a richer variety of indoor objects and exhibits complex, fine-grained spatial-physical relationships. In contrast, the sample from 3D-FRONT (right) lacks such fine-grained object interactions, particularly with small objects, highlighting the relational sparsity inherent in the original dataset that our preprocessing pipeline aims to address}
        \label{fig:comparison_dataset}
\end{figure*}

While existing 3D scene datasets such as 3D-FRONT and ProcTHOR have made significant strides in scale and quality, they share a fundamental limitation: a pronounced lack of rich, fine-grained object-to-object relations. The 3D-FRONT dataset primarily consists of designer-curated indoor scenes, where layouts are often aesthetically oriented and orderly. This results in oversimplified spatial relationships (e.g., "against a wall" or "centered in a room") and fails to capture the complex, unstructured interactions commonly found in real-world environments, such as object stacking, partial occlusion, or casual support. Although the ProcTHOR dataset offers greater environmental diversity through procedural generation and includes more small-scale objects, its automated process does not explicitly or comprehensively annotate crucial physical support relations or precise spatial relations. This inherent sparsity of relational data makes it difficult for models trained directly on these datasets to learn and reason about the complex spatial and physical interactions characteristic of real-world scenes. Consequently, models frequently exhibit fine-grained physical inconsistencies, such as floating objects, inter-penetrations, and implausible support structures. This critical shortcoming motivates our work to not only rely on raw data but also to introduce a comprehensive preprocessing pipeline (detailed in Section 7.2) that explicitly derives these missing relations, thereby compensating for the inherent limitations of the source data and providing a foundation for learning robust spatial-physical reasoning. Here we present the comparison of our preprocessed dataset and the original 3D-FRONT dataset in Fig.\ref{fig:comparison_dataset}.

\subsection{Preprocessing}
We conduct experiments on the \textbf{ProcTHOR-10K} dataset, a large-scale collection of 10,000 procedurally generated 3D indoor environments developed by the Allen Institute for AI. We implement a comprehensive, parallelized preprocessing pipeline using Blender for geometric operations.

The pipeline executes the following automated steps for each scene. First, the full house layout is deconstructed into individual rooms based on architectural boundaries. Within each room, non-supporting objects such as wall art and small decorative items are filtered out, while functional furniture such as tables, shelves, and countertops are retained. To ensure physical plausibility, the pipeline automatically rectifies geometric issues, including resolving mesh intersections between objects and grounding floating objects onto their nearest underlying supporting surfaces. Subsequently, supported objects are identified and paired with their corresponding supporting objects (e.g., a \texttt{Mug} on a \texttt{Table}). Relative spatial relationships (e.g., \texttt{left}, \texttt{right}) are computed and annotated for all object pairs.

Finally, the structured data, including object poses, meshes, and their annotated relationships, is serialized into \texttt{.npz} format for efficient loading during training.

\begin{table}[h]
  \caption{\textbf{Dataset Statistics.} Summary of the processed ProcTHOR-10K dataset, detailing the scale of rooms, splits, and object diversity.}
  \centering
  \footnotesize
  \begin{tabular}{lc}
    \toprule
    Metric & Value \\
    \midrule
    Total Processed Rooms & 23,472 \\
    Training Set Size (rooms) & 22,472 \\
    Test Set Size (rooms) & 1,000 \\
    Unique Object Categories & 108 \\
    Objects per Room (range) & 2 - 52 \\
    \bottomrule
  \end{tabular}
  \label{tab:dataset_stats}
\end{table}

\section{Explanation for metrics}

To provide a multifaceted evaluation of our method, we employ a suite of metrics assessing visual fidelity, physical plausibility, and relational accuracy.

\noindent\textbf{Fréchet Inception Distance (FID$\downarrow$)}. We assess the visual quality and diversity of our generated scenes by rendering top-down 2D views and calculating the FID score against rendered images from the ground-truth test set. This measures the distributional similarity between generated and real scene layouts. However, this metric is limited in its applicability to our method, as our approach focuses more on physical plausibility, which the FID score cannot reflect.

\noindent \textbf{Mesh Collision Rate ($\text{Col}_{\text{mesh}}$$\downarrow$)}. This metric quantifies the degree of inter-penetration between objects. For each scene, we use \texttt{trimesh.CollisionManager} to identify all intersecting mesh pairs. We consider a collision significant if the penetration depth exceeds a threshold of 0.01 units, ignoring minor surface contacts. The final score reports the percentage of objects involved in at least one significant collision across the entire test set. A lower score indicates superior physical plausibility.

\noindent \textbf{Graph Recall (GRecall$\uparrow$)}. To measure structural accuracy, we evaluate the adherence of generated scenes to the input spatial-relation graph. For each generated scene, we derive the pairwise relative spatial relationships between all objects based on their final poses. GRecall measures the proportion of relationships in the ground-truth input graph that are correctly realized in the generated scene.

\noindent\textbf{Average Support Distance (ASD$\downarrow$)}. This metric evaluates the physical quality of support relationships. For each supported object, we compute the signed distances from its vertices to the supporting object. The absolute value of the minimum signed distance---representing the minimum gap or largest penetration---is taken as the support distance. ASD is the average of these distances over all support pairs. A lower value signifies more precise, high-quality contact.

\noindent\textbf{Isaac Stability (Stability$\uparrow$)}. To rigorously test physical robustness, we perform ten 100-step physics simulations for each generated scene in NVIDIA Isaac Sim. We record the poses of all objects before and after the simulation. Stability is measured as the percentage of pairwise spatial relationships that remain unchanged after the simulation concludes. A high stability score indicates that the generated layout is physically sound and resilient to gravity-induced collapse.

\section{Ablation studies}
In the main paper, we discussed the effectiveness of each design in \ours{}, covering model architecture and guidance choice. Here, we present further methodological details and ablation analysis.

\noindent \textbf{What is the effect of choosing Michelangelo encoder?} While conventional encoders, such as those based on open-shape or trained on small datasets, can characterize shape, they often only offer simple image-text alignment semantic features or object identifiers specific to their small dataset. When selecting a shape encoder to extract geometric features, we consider not only the compatibility between the shape embedding and the relational input embeddings but also its capacity for capturing complex underlying geometric information. The heightened requirements prompt us to employ a special encoder capable of both aligning the shape latent space with semantic modalities and capturing underlying geometric information. Our early experiments revealed modest improvement compared with the plain shape encoder in performance metrics like the rate of mesh collision in a scene.

\noindent \textbf{What is the effect of geometry-aware module in scene diffusion?} In this paper, we posit that geometry-aware diffusion is central to enabling generative models to understand and generate spatially and physically plausible scenes. Without the explicit integration of geometric information, conventional scene generation merely learns the statistical distribution of the dataset, a process inherently limited by the dataset's scale and quality. However, the dataset distribution is insufficient to capture complex inter-object relationships and therefore fails to address object intersections within a scene. This relatively unexplored problem motivates us to integrate explicit awareness of the geometric state into the scene generation process at the current timestep. To evaluate our method's performance without geometry feature diffusion, we remove the geometry-aware module. Consequently, the ablated model produces scenes with higher colliding rates and floating artifacts. This validates the effectiveness of geometry-aware module in achieving understanding and generating spatial and physically plausible results.

\noindent \textbf{What is the effect of perceiver-based transformer as geometry-aware module?} We advocate the use of perceiver-based transformer as our geometry-aware network. The architecture allows for processing large-scale point cloud inputs without structural modification, effectively reducing complexity of model without compromising performance. An alternative is to employ plain transformer. However, we observe it achieves equal performance at higher computational cost and deeper network structure.

\section{Baselines}
\noindent \textbf{ATISS} \cite{paschalidou2021atiss} generates indoor scenes by autoregressively adding objects. At each step, the model encodes the set of previously placed objects. It subsequently predicts the attributes including 3D location, size, orientation, and semantic category. 

\noindent \textbf{DiffuScene}. DiffuScene represents each scene as a fixed-size set of objects, where each object is characterized by a concatenation of properties including 3D location, size, orientation, semantic category, and geometric feature. During the generation, the model iteratively denoises a randomly initialized set of object attributes using a UNet-1D architecture with attention mechanisms. After denoising, object geometries are retrieved from a 3D database using the predicted shape codes and semantic labels.

\noindent \textbf{InstructScene}. InstructScene synthesizes 3D indoor scenes through a two-stage generative framework. In the first stage, the model learns a semantic graph prior capturing high-level object relationships and appearance features conditioned on natural language instructions. In the second stage, a layout decoder generates precise 3D layout attributes-including location, size, and orientation—for each object by denoising a noisy latent representation derived from the semantic graph. However, it relies on object retrieval from a predefined asset library, which inherently introduces limitations.

\section{Additional results}
In Fig.8, we provide additional qualitative results and zoom-in visualization, demonstrating our work's capability to generate physically plausible and meaningful relational results.

\section{User study}

We conducted a perceptual user study to evaluate the physical and semantic plausibility of our method in comparison to ATISS, DiffuScene and InstructScene on the scene generation task. Fig.\ref{fig:user_supp} displays the scenes generated by the four methods alongside their corresponding post-physics simulation states. For each pair of results, participants were asked to select the scene exhibiting greater physical consistency and scene rationality. We collected 57 valid responses and calculate the statistics. Our method secured 88.6\% of the votes, demonstrating a substantial advantage over ATISS (0.9\%), InstructScene (4.4\%), and DiffuScene (6.1\%).

\begin{figure*}[htbp]
    \centering
    \begin{minipage}[t][0.95\textheight]{\textwidth}
        \centering
        
        \begin{minipage}[b]{0.495\textwidth}
            \includegraphics[width=\textwidth]{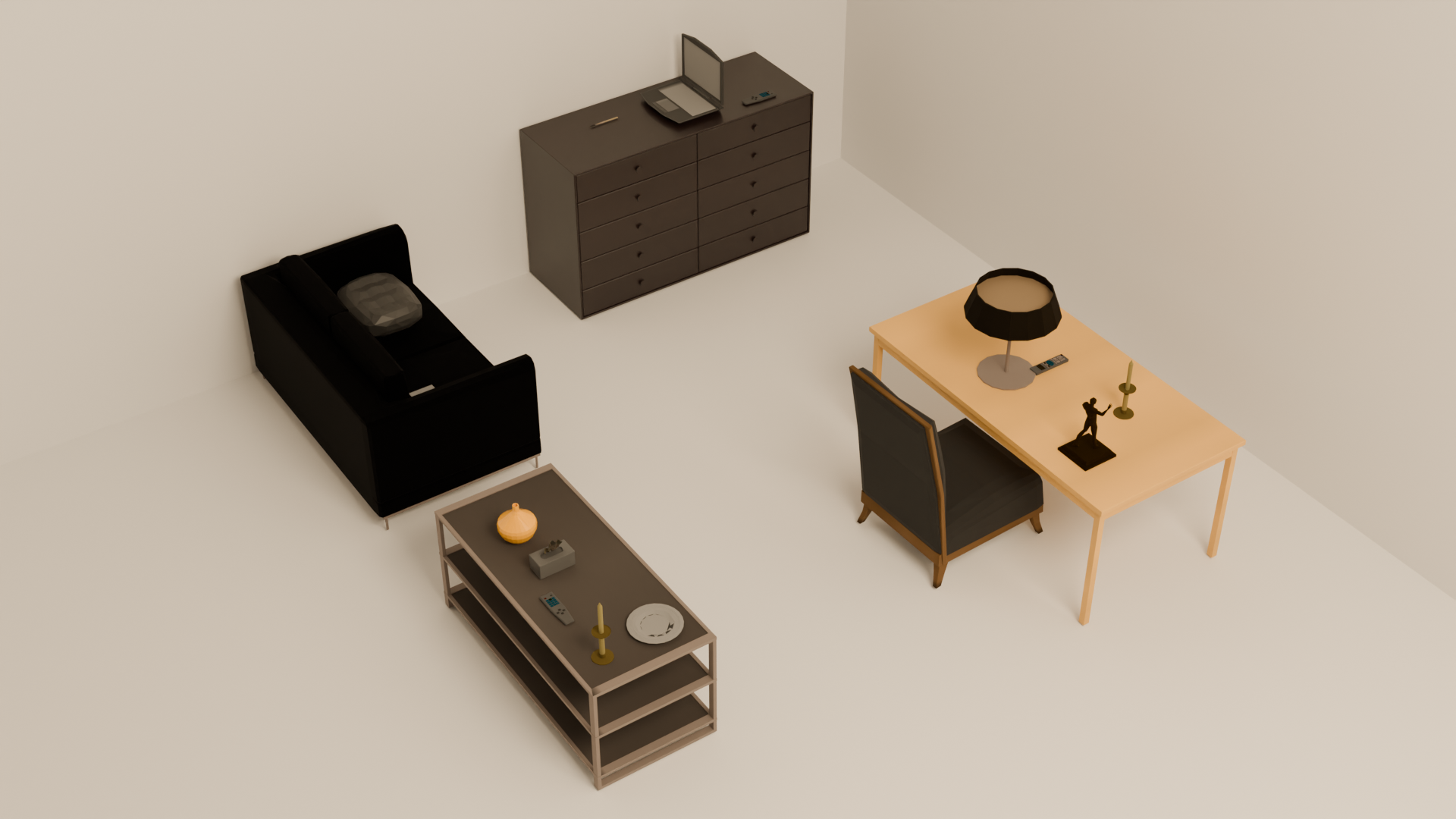}
        \end{minipage}
        \hfill
        \begin{minipage}[b]{0.495\textwidth}
            \includegraphics[width=\textwidth]{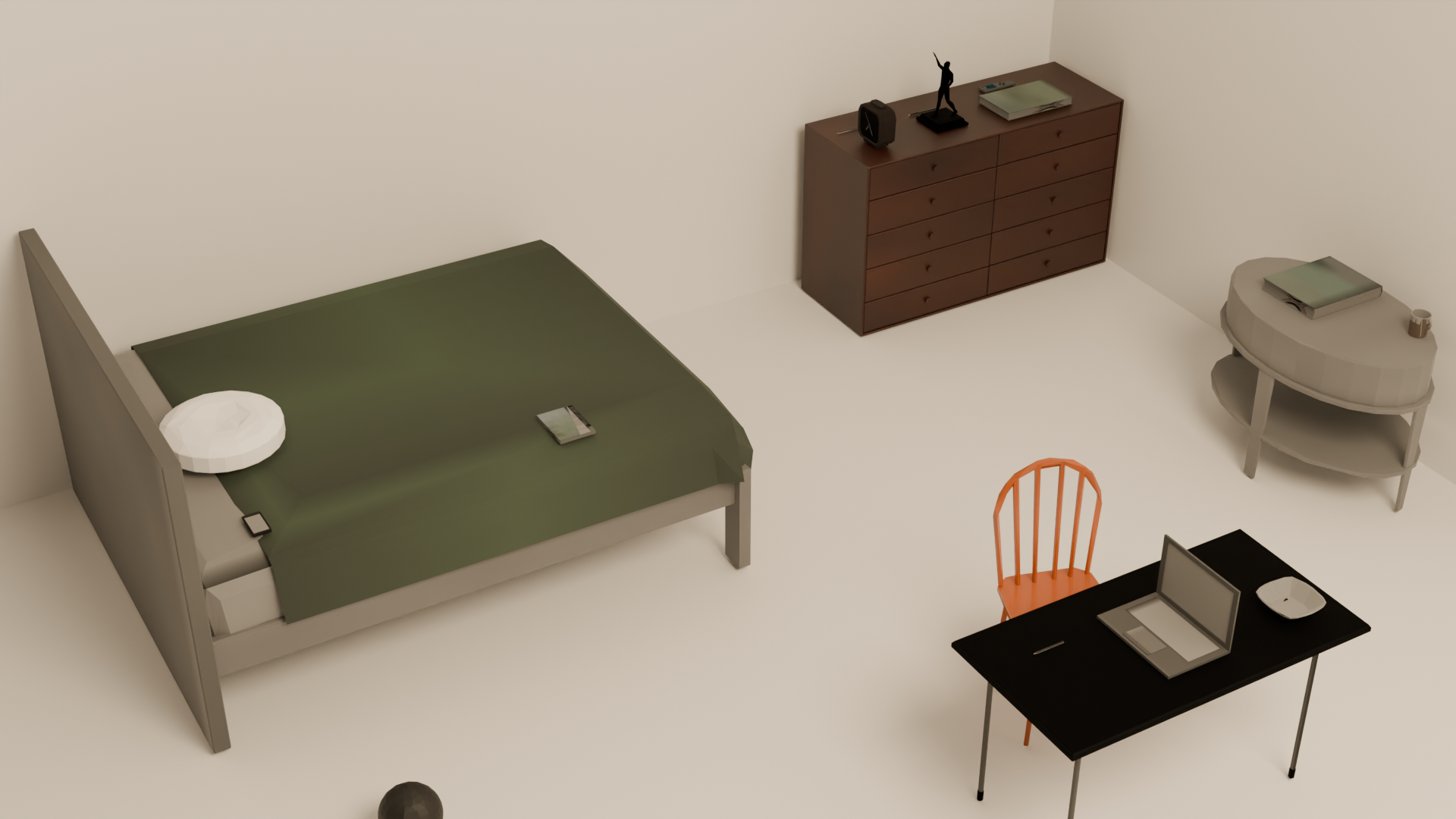}
        \end{minipage}
        
        \vfill
        
        \begin{minipage}[b]{0.495\textwidth}
              \includegraphics[width=\textwidth]{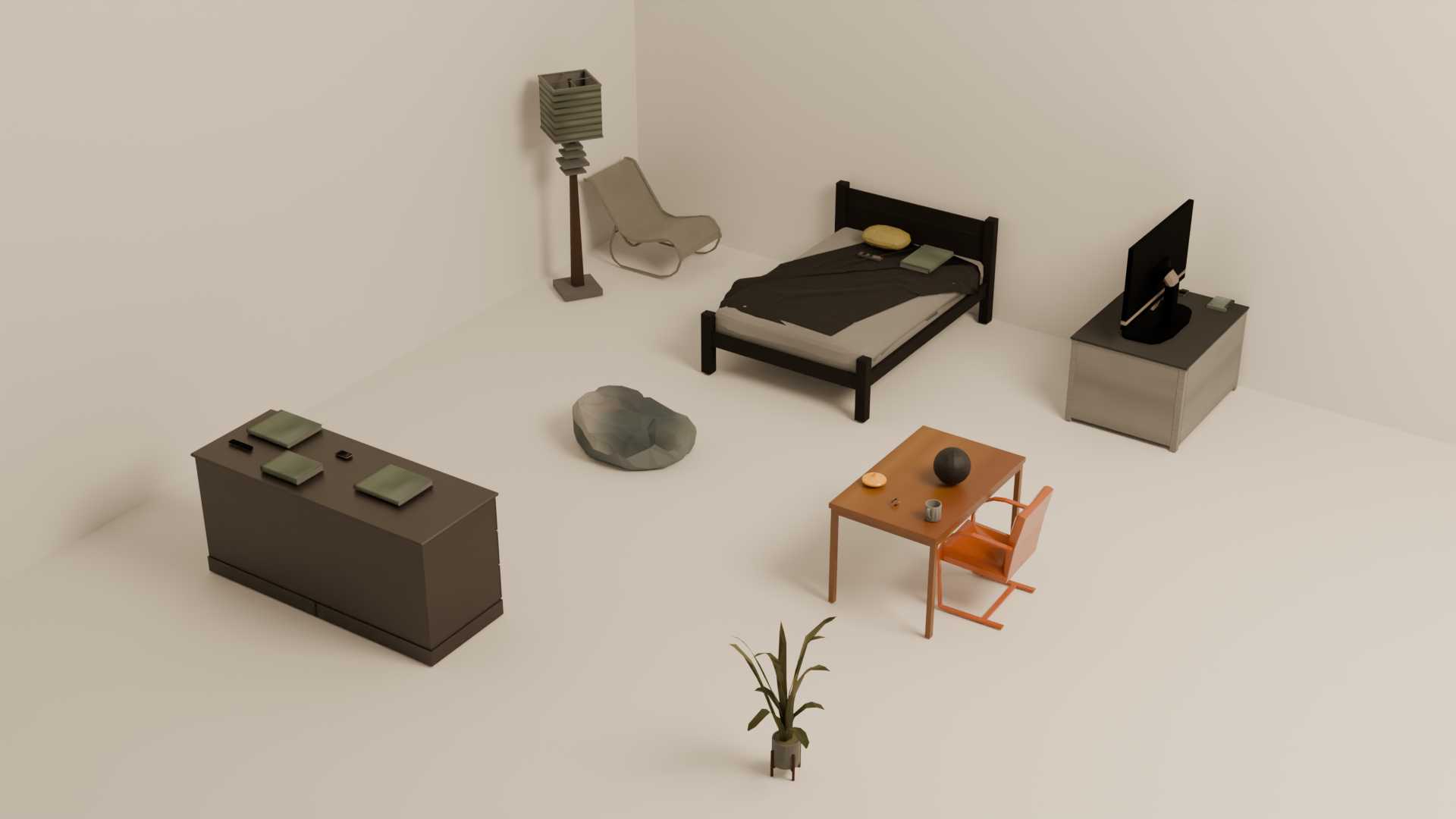}
        \end{minipage}
        \hfill
        \begin{minipage}[b]{0.495\textwidth}
            \includegraphics[width=\textwidth]{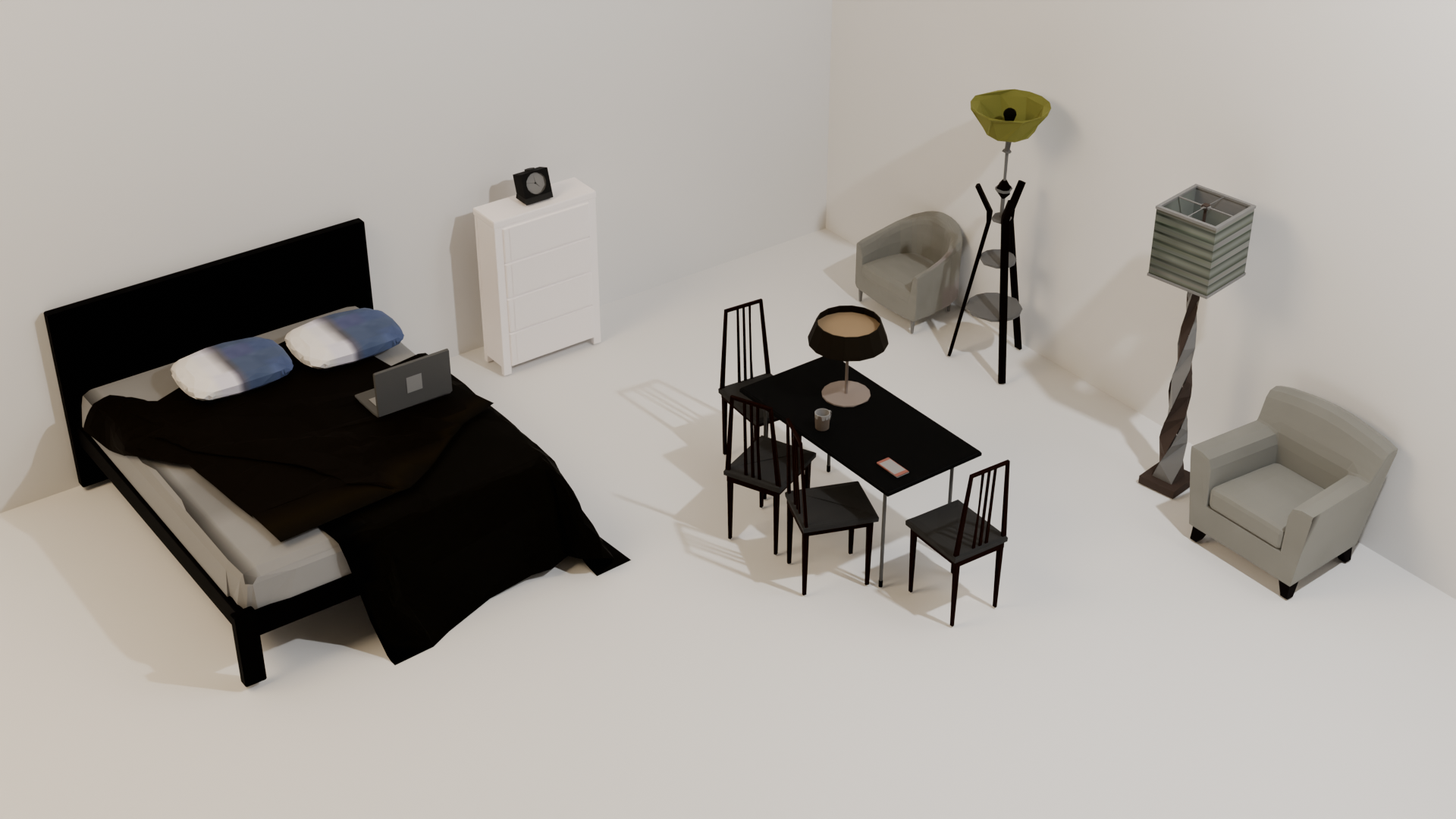}
        \end{minipage}
        
        \vfill
        
        \begin{minipage}[b]{0.495\textwidth}
            \includegraphics[width=\textwidth]{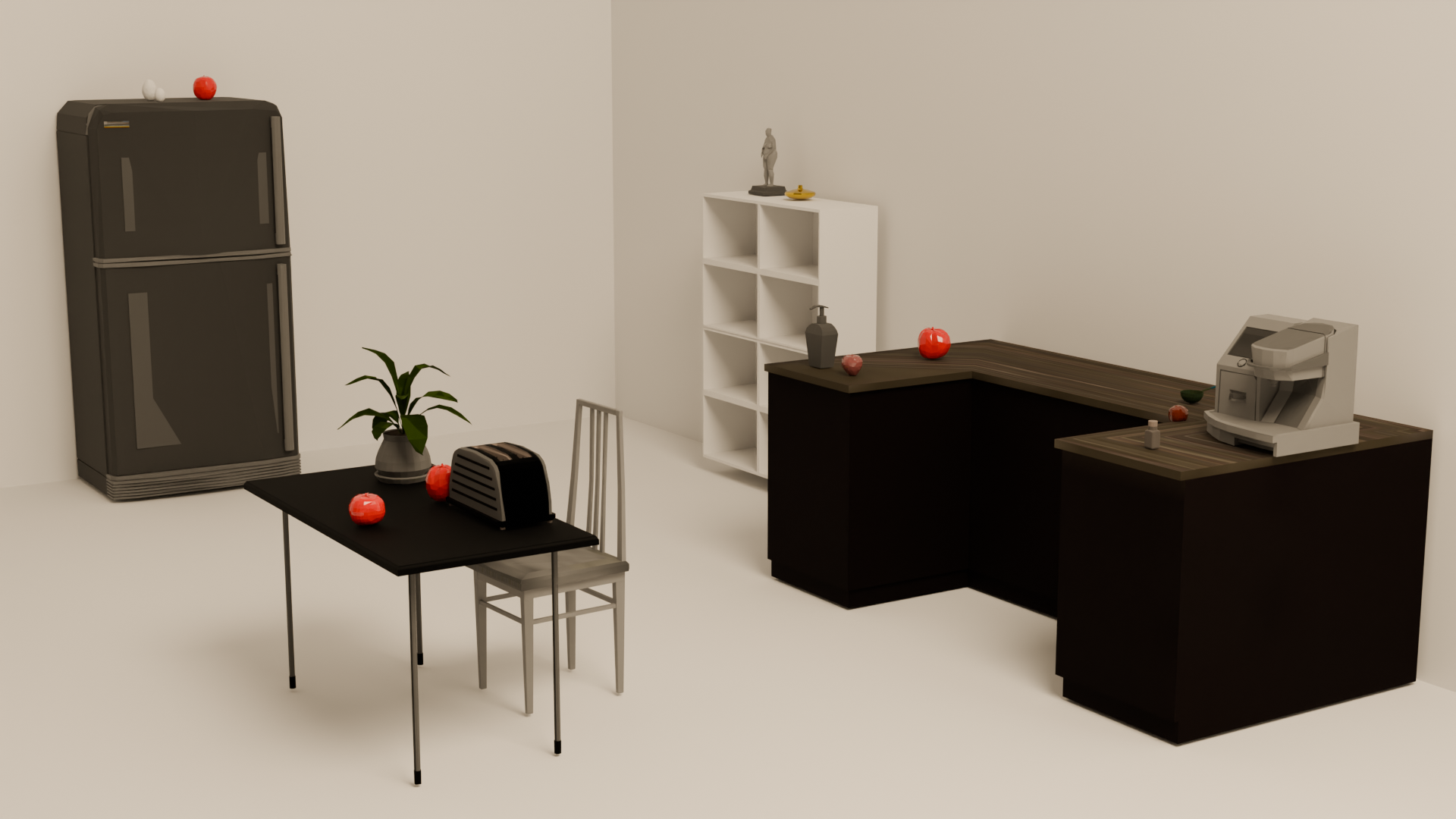}
        \end{minipage}
        \hfill
        \begin{minipage}[b]{0.495\textwidth}
            \includegraphics[width=\textwidth]{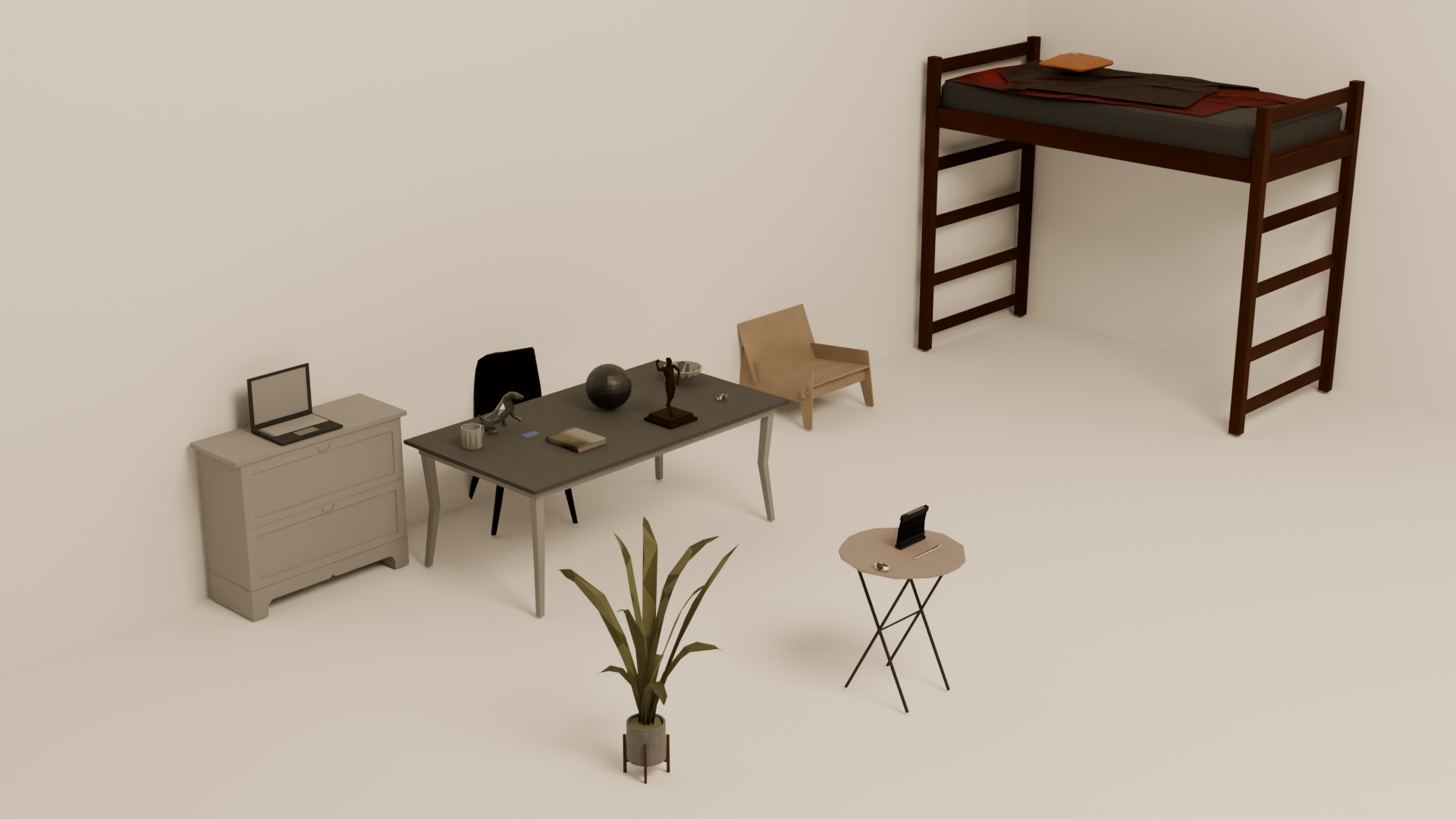}
        \end{minipage}
        
        \vfill
        
        \begin{minipage}[b]{0.495\textwidth}
            \includegraphics[width=\textwidth]{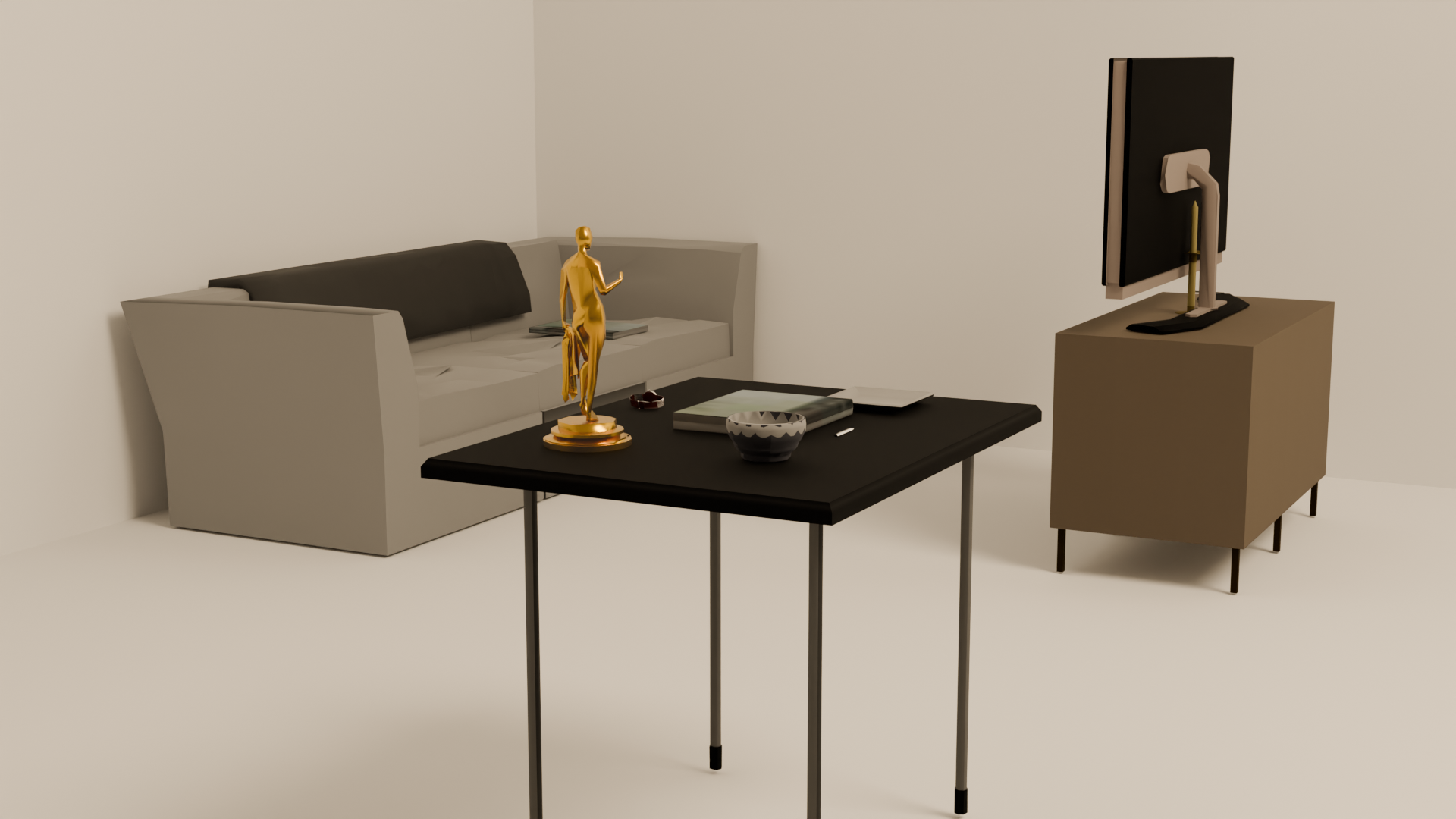}
        \end{minipage}
        \hfill
        \begin{minipage}[b]{0.495\textwidth}
            \includegraphics[width=\textwidth]{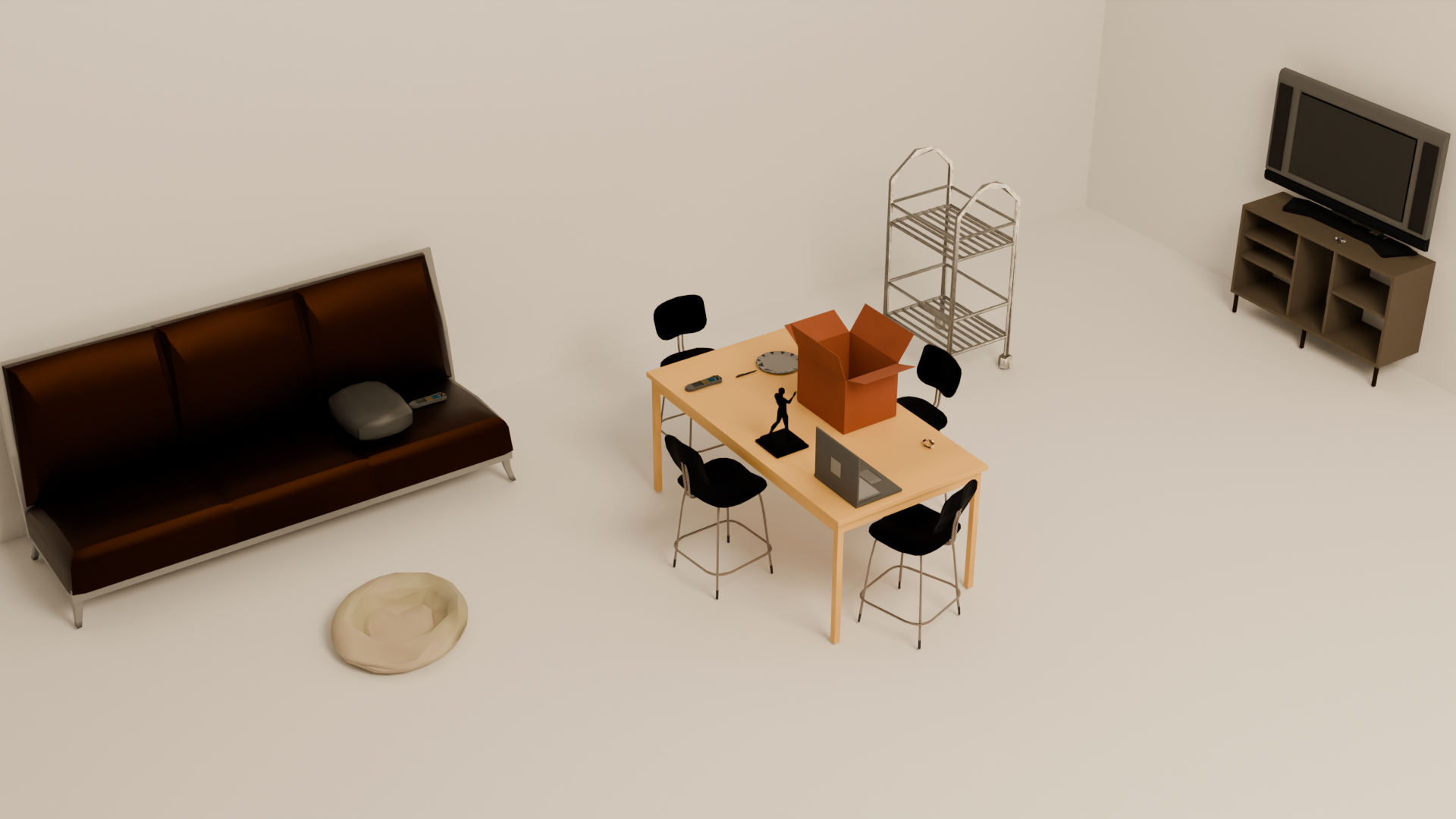}
        \end{minipage}
        
        \vfill
        
        \caption{\textbf{Additional Qualitative results}. The gallery displays 8 randomly selected samples, demonstrating the diversity and physical plausibility of the generated 3D scenes.}
    \end{minipage}
    \label{fig:full_page_gallery}
\end{figure*}

\begin{figure*}[!t]
  \centering
  \includegraphics[width=\textwidth]{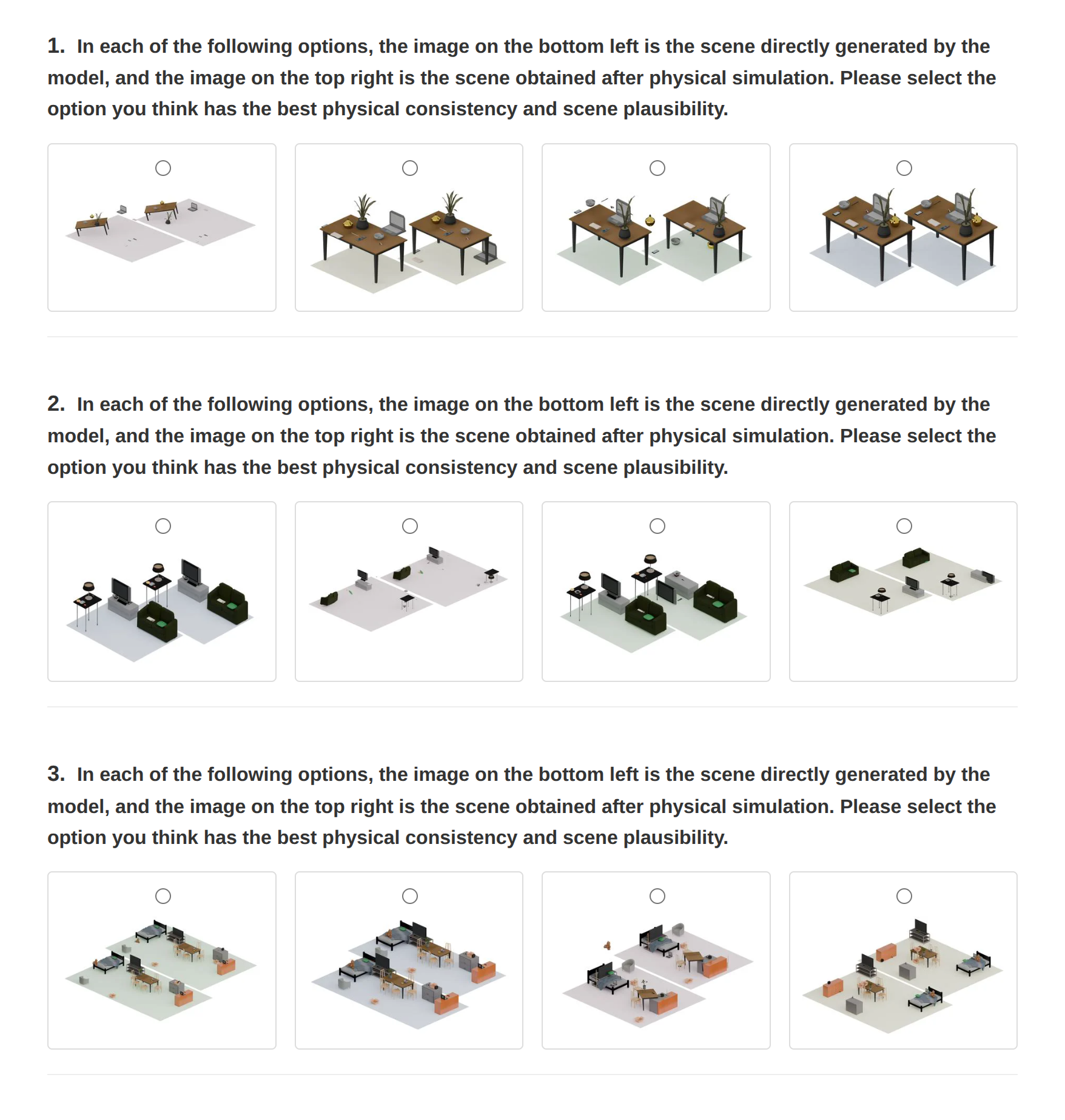}
  \caption{\textbf{User Study UI.} Based on the generated scene and the scene after physical simulation, which were available for detailed inspection via a zoom function, participants were asked to choose the pair exhibiting greater physical consistency and scene rationality.}
  \label{fig:user_supp}
\end{figure*}